\documentclass[prd,aps,showpacs,floats,floatfix,superscriptaddress,nofootinbib]{revtex4}

\usepackage{graphicx}
\usepackage{amsmath}
\usepackage{amssymb}
\usepackage{amsfonts}
\usepackage{bbm}
\usepackage{amscd}
\usepackage{array}
\usepackage{tensind}
\usepackage{mathrsfs}
\tensordelimiter{?}

\usepackage{longtable}

\def\laq{\raise 0.4ex\hbox{$<$}\kern -0.8em\lower 0.62ex\hbox{$\sim$}}
\def\gaq{\raise 0.4ex\hbox{$>$}\kern -0.7em\lower 0.62ex\hbox{$\sim$}}

\newcommand{\beq}{\begin{equation}}
\newcommand{\eeq}{\end{equation}}
\newcommand{\bea}{\begin{eqnarray}} 
\newcommand{\eea}{\end{eqnarray}}
\newcommand{\ba}{\begin{array}}
\newcommand{\ea}{\end{array}}

\def\vct#1{{\mathchoice{\mbox{\boldmath$#1$}}{\mbox{\boldmath$#1$}}%
  {\mbox{\scriptsize\boldmath$#1$}}{\mbox{\scriptsize\boldmath$#1$}}}}

\newcommand{\mytextrm}[1]{{}}

\def\fracpd#1#2{\frac{\partial#1}{\partial#2}}

\def\ee{{\mathrm{e}}}

\def\dd{{\mathrm{d}}}

\newlength{\sizeonefig}
\newlength{\sizetwofig}
\newlength{\sizeonefigb}
\newlength{\sizetwofigb}
\begin{document}

\title{Higher-order-in-spin interaction Hamiltonians for binary black holes from source terms of Kerr geometry in approximate ADM coordinates}

\author{Steven Hergt} 

\author{Gerhard Sch\"afer} 

\affiliation{Friedrich-Schiller-Universit\"at Jena, Max-Wien-Platz 1, 07743 Jena, Germany}

\begin{abstract}
The Kerr metric outside the ergosphere is transformed into Arnowitt-Deser-Misner coordinates up
to the orders $1/r^4$ and $a^2$, respectively in radial coordinate $r$
and reduced angular momentum variable $a$, starting from the Kerr
solution in quasi-isotropic as well as harmonic coordinates. The distributional source terms for the
approximate solution are calculated. To leading order in linear momenta, higher-order-in-spin interaction
Hamiltonians for black-hole binaries are derived.
\end{abstract}


\date{\today}

\maketitle

\section{Introduction}
\label{sec0}
The motion of binary compact objects with proper rotations is an
important topic in general relativity. Very likely, the first detection
of gravitational waves will originate from those sources. Recent
achievements of the numerical relativity community in the
simulation of merging spinning binary black holes demand deeper analytic work in the
treatment of the motion of binary black holes to compare with and
hopefully to support the obtained results, \cite{RDR, BKN, MTB}.

In the analytic treatment of the motion of compact binaries with
point-like components the Arnowitt-Deser-Misner (ADM) canonical formalism has proved very
efficient, \cite{DJS01, JS98}; also see \cite{schaefer2005}.
It is thus highly desirable to generalize that formalism to spinning
compact binaries. The aim of this paper is to transfer information
from spinning black holes at rest, i.e. from Kerr black holes \cite{kerr}, onto the dynamics
of binary black holes with spin. For reaching this goal, at
first the Kerr metric is approximately transformed into coordinates which are crucial for the
ADM canonical formalism. Then the source terms connected with this
form of the metric are computed. Finally, new terms for the binary
interaction Hamiltonian are constructed. 

The paper heavily relies on the diploma thesis by one of the authors, \cite{hergt}. 
In the paper, the theory of generalized functions is extensively used, \cite{gelfand}.
Throughout the paper, the speed of light $c$  and the Newtonian
gravitational constant $G$ are put equal to 1. Greek indices will run
through 0,1,2,3; Latin indices through 1,2,3. The signature of the metric is +2.

\section{(3+1)-splitting and ADM formalism}

In this section the ADM formalism is presented as much as it is needed for
the purpose of our paper, \cite{ADM}. 

The $(3+1)$-decomposition \cite{gravitation} of a spacetime manifold gives rise to the following line element:

\begin{equation}\label{lineelement}
\dd s^2=(N_{i}N^i-N^2)\dd t^2+2N_{i}\dd x^i\,\dd t+\gamma_{ij}\dd
x^i\,\dd x^j=-N^2\dd t^2+\gamma_{ij}(\dd x^i+N^i \dd t)\,(\dd x^j+N^j\dd t)\,,
\end{equation}
where  $N$ is the lapse function, $N^{i}$ ($N_i = \gamma_{ij}N^j$) the shift vector, and 
$\gamma_{ij}$ the $3$-metric of the $x^0=t=\text{const.}$ spacelike hypersurfaces. The following
relations between these quantities and the metric tensor hold,
\begin{equation}\label{downmetric}
g_{\mu\nu}=
\begin{pmatrix}
-N^2+N_{i}N^{i} & N_{i}\\
N_{j} & ?\gamma_{ij}?
\end{pmatrix}\,,
\end{equation}
\begin{equation}
?g^{\mu\nu}?=
\begin{pmatrix}
-1/N^{2} & N^{i}/N^2\\
N^{j}/N^2 & \gamma^{ij}-\frac{N^{i}N^{j}}{N^2}
\end{pmatrix}
\end{equation}
with $\gamma^{ij}$ being the inverse metric of $\gamma_{ij}$, $?\gamma^{ik}??\gamma_{kj}?=\delta_{ij}$.\\

The metric tensor (\ref{downmetric}) admits a gauge transformation via a diffeomorphism $f$ resulting in a transformation formula

\begin{equation}\label{metrictrafo}
g^{N}_{\mu\nu}(x):=g^{O}_{\alpha\beta}(f(x))\fracpd{f^{\alpha}}{x^{\mu}}\fracpd{f^{\beta}}{x^{\nu}}\,,
\end{equation}
$\lbrace f^{\alpha}\rbrace$ being the old coordinates depending on the new ones $\lbrace x^{\mu}\rbrace$.\\

The projections of the energy-momentum tensor $T_{\mu\nu}$ related with
the $(3+1)$-slicing of spacetime read, see, e.g., \cite{durancfc+}
\begin{align}
E&\equiv n^{\mu}n^{\nu}T_{\mu\nu}=N^2T^{00}\,,\label{Egleichung}\\
S_{i}&\equiv-\bot^{\mu}_{i}n^{\nu}T_{\mu\nu}=-\frac{1}{N}\left(T_{0i}-T_{ij}N^{j}\right)= N T^0_i\,,\label{Sigleichung}\\
S_{ij}&\equiv\bot^{\mu}_{i}\bot^{\nu}_{j}T_{\mu\nu}=T_{ij}\,,
\label{Sijgleichung}
\end{align}
where $n_{\mu} = (-N,0,0,0)$ is the unit vector orthogonal to the
$t=$ const. slices and $ \bot^{\mu}_{i}=\delta^{\mu}_i$ denotes the
projection tensor onto these slices. The source terms in the ADM
formalism read \cite{ADM},    
\begin{align}
E^{*}&\equiv\sqrt{\gamma}E=N\sqrt{-g}T^{00}\,,\label{estab}\\
S^{*i}&\equiv\sqrt{\gamma}S^{i}=\gamma^{ij}\sqrt{-g}T^{0}_{j}\,,\label{sstab}\\
S^{*ij}&\equiv\frac{1}{2}N\sqrt{\gamma} S^{ij} =\frac{1}{2}\gamma^{ki}\gamma^{lj}\sqrt{-g}T_{kl} \label{ssstab}
\end{align}
and appear in the field equations as follows, 

\begin{equation}
-\frac{1}{\sqrt{\gamma}}\left(\pi^{ij}\pi_{ij}-\frac{1}{2}\pi^{2}\right)+\sqrt{\gamma}R_{(3)}=16\pi
E^{*}  \quad \quad \mbox{(energy constraint equation)}\,,
\end{equation}
\begin{equation}
-?\pi^{ik}_{;\,k}?=8\pi S^{*i} \quad \quad \mbox{(momentum constraint equations)}\,,
\end{equation}
\begin{equation}
\partial_{t}\gamma_{ij}= -2N\gamma^{-1/2}(\pi_{ij} -\frac{1}{2}\pi\gamma_{ij})
+ N_{i;j} + N_{j;i} \quad \quad \mbox{(evolution equations)}\,,
\end{equation}
\begin{equation}
\begin{split}
\partial_{t}\pi^{ij}=&\,-N\sqrt{\gamma}\,(^{(3)}R_{ij}-\frac{1}{2}\gamma^{ij}\,R_{(3)})+\frac{1}{2}N\gamma^{-1/2}\gamma^{ij}(\pi^{mn}\pi_{mn}-\frac{1}{2}\pi^{2})-2N\gamma^{-1/2}(\pi^{im}?\pi_{m}^{j}?-\frac{1}{2}\pi\pi^{ij})\\
&\,+\sqrt{\gamma}(N^{;ij}-\gamma^{ij}?N^{;m}_{;m}?)+(\pi^{ij}N^{m})_{;m}-?N^{i}_{;m}?\pi^{mj}-?N^{j}_{;m}?\pi^{mi}
+ 16\pi S^{*ij} \quad \quad \mbox{(evolution equations)}\,,
\end{split}
\end{equation}
where $\pi^{ij}_{true}$ with 
\begin{equation}\label{momentum}
16\pi\pi^{ij}_{true} \equiv \pi^{ij}=\sqrt{-g}\left(\Gamma^{0}_{pq}-\gamma_{pq}\Gamma^{0}_{rs}\gamma^{rs}\right)\gamma^{ip}\gamma^{jq}
\end{equation}
is the canonical momentum density of the gravitational field, $\pi =
\gamma_{ij}\pi^{ij}$ and ``;" denotes the 3-dimensional covariant derivative. 

The ADM Hamiltonian results from 
\begin{equation}\label{ham01}
H_{\text{ADM}}=\frac{1}{16\pi}\oint_{\infty}\dd^2 s_{i}(\gamma_{ij,\,j}-\gamma_{jj,\,i})
\end{equation}
after insertion of a solution of the constraint equations subjected to 
appropriate coordinate conditions.

\subsection{ADMTT coordinates}

To get a unique solution of the constraint and evolution equations the
coordinates have to be fixed by a gauging process. The ADM transverse traceless (ADMTT) gauge
refers to generalized isotropic coordinates defined by the conditions \cite{ADM}

\begin{equation}{\label{ADM}}
\gamma_{ij}=\left(1+\frac{1}{8}\phi\right)^{4}\delta_{ij}+h_{ij}^{TT}\;, \quad \pi^{ii}=0
\end{equation}
with $h_{ij}^{TT}$ being transversal and tracefree ($h_{ij,j}^{TT}=0$\;, $h_{ii}^{TT}=0$).
The TT-part of a (symmetric) tensor of rank two can be achieved by application of the
TT-projection operator $\delta^{TTkl}_{ij}$ defined by
{\setlength\arraycolsep{2pt}
\begin{equation}
\begin{array}{rcl}
\displaystyle
\delta^{TTkl}_{ij}&:=&\frac{1}{2}\bigg[(\delta_{il}-\Delta^{-1}\partial_{i}\partial_{l})(\delta_{jk}-\Delta^{-1}\partial_{j}\partial_{k})+
(\delta_{ik}-\Delta^{-1}\partial_{i}\partial_{k})(\delta_{jl}-\Delta^{-1}\partial_{j}\partial_{l})\\[8pt]
& & \quad-(\delta_{kl}-\Delta^{-1}\partial_{k}\partial_{l})(\delta_{ij}-\Delta^{-1}\partial_{i}\partial_{j})\bigg]\,.
\end{array}
\end{equation}}

\subsection{The conformal flatness condition}

Under the conformal flatness condition (CFC) for the 3-metric
$\gamma_{ij}$, 

\begin{equation}
\gamma_{ij}^{CFC}=\psi^{4}\delta_{ij}\,, \quad \psi=1+\frac{1}{8}\phi\,,
\end{equation}
which on account of $\pi^{ii} =0$ results in $\pi =0$ (maximal slicing), 
the equations for $\psi$ $N$, and $N^{i}$ read, see, e.g. \cite{durancfc+},

\begin{align}
\Delta\psi=&-2\pi\left(\psi^{-1}E^{*}+\psi^{-7}\frac{\hat{K}_{ij}\hat{K}^{ij}}{16\pi}\right)\,,\label{1kon}\\
\Delta(N\psi)=&2\pi N\left[\psi^{-1}(E^{*}+2S^{*})+\psi^{-7}\frac{7\hat{K}_{ij}\hat{K}^{ij}}{16\pi}\right]\,,\label{2lapse}\\
\Delta N^{i}=&16\pi N\psi^{-2}S^{*i}+2\hat{K}^{ij}\partial_{j}\left(\frac{N}{\psi^{6}}\right)-\frac{1}{3}\partial^{i}\partial_{k}N^{k}\,,\label{3shift}
\end{align}
where $S^{*} \equiv \sqrt{\gamma} \gamma_{ij}S^{ij}$ and where $\hat{K}_{ij}$ is defined by
\begin{equation}\label{flkrum}
\hat{K}_{ij}=\frac{1}{2N}\left(\partial_iN^{j}+\partial_jN^{i}-\frac{2}{3}\delta_{ij}\partial_kN^{k}\right)\,,
\end{equation}
(notice here: ${\hat{K}}_{ij}{\hat{K}}^{ij}=\pi_{ij}\pi^{ij}$).

\subsection{CFC+ approximation}

In regard to our Kerr metric calculations, the CFC metric will be
generalized to the leading order deviation (LO) term from isotropy,
\cite{durancfc+}

\begin{equation}
\gamma_{ij}^{CFC+}=\psi^{4}\delta_{ij}+h_{ij}^{\text{LO}}\,.
\end{equation}

In the ADMTT gauge $h_{ij}^{\text{LO}}$ turns out to be the leading
order part of $h_{ij}^{TT}$. This $h_{ij}^{TT}$ is a solution of the
tensor Poisson equation

\begin{equation}
\Delta h_{ij}^{TT}=\delta^{TTkl}_{ij}\,F_{kl}\,,
\end{equation}
where the source $F_{kl}$ is given by

\begin{equation}
F_{kl}=-4U_{,\,k}U_{,\,l}-16\pi\frac{S_{k}^{*}S_{l}^{*}}{E^{*}}
\end{equation}
with $U$ being the ``Newtonian" potential which is a solution of the Poisson equation
\begin{equation}
\Delta U=-4\pi E^{*}.
\end{equation}

Considering the CFC equations above for $N$, $\psi$, and $N^{i}$ only the equation for the lapse function has to be modified:
\begin{equation}
\Delta(N\psi)=\left[\Delta(N\psi)\right]_{h_{ij}^{TT}=0}-\delta^{ik}\delta^{jl}h_{ij}^{TT}U_{,\,kl}.
\end{equation}
However, this term will turn out be to neglectible for the Kerr metric
in our approximation.

\section{Approximate ADM coordinates for Kerr metric}
\label{sec2}

\subsection{Transformation from quasi-isotropic to ADMTT coordinates}

We start with the line element in quasi-isotropic coordinates \cite{brandt, cookinitial}

\begin{equation}\label{quasi}
\dd s^2=-N^2\dd t^2+\psi^4\left[\ee^{2\mu/3}(\dd\tilde{r}^2+\tilde{r}^2\dd\theta^2)+\tilde{r}^2\sin^2\theta\,\ee^{-4\mu/3}(\dd\phi+N^{\phi}\dd t)^2\right]
\end{equation}
with
\begin{eqnarray}
N^2&=&\frac{\rho^2\Delta}{(r^2+a^2)^2-\Delta a^2\sin^2\theta}\,,\label{lapseq1}\\
N^{\phi}&=&-a\frac{r^2+a^2-\Delta}{(r^2+a^2)^2-\Delta a^2\sin^2\theta}\,,\\
\psi^4&=&\frac{\rho^{2/3}\left((r^2+a^2)^2-\Delta a^2\sin^2\theta\right)^{1/3}}{\tilde{r}^2}\,,\\
\ee^{2\mu}&=&\frac{\rho^4}{(r^2+a^2)^2-\Delta a^2\sin^2\theta}
\end{eqnarray}
and
\begin{eqnarray}
\rho^2&=&r^2+a^2\cos^2\theta,\qquad\Delta=r^2-2mr+a^2\,,\\
r&=&\tilde{r}\left(1+\frac{m+a}{2\tilde{r}}\right)\left(1+\frac{m-a}{2\tilde{r}}\right)
\label{schlange}
\end{eqnarray}
with $(r,\,\theta,\,\phi)$ denoting the usual
Boyer-Lindquist coordinates. Notice that in the limit $a\rightarrow 0$
we get back the Schwarzschild
metric in spatial isotropic coordinates
\begin{equation}\label{schwarziso}
\dd s^2=-\left(\frac{1-\frac{m}{2\tilde{r}}}{1+\frac{m}{2\tilde{r}}}\right)^2\dd t^2+\left(1+\frac{m}{2\tilde{r}}\right)^4(\dd\tilde{r}^2+\tilde{r}^2\dd\theta^2+\tilde{r}^2\sin^2\theta\dd\phi^2)\,.
\end{equation}

We write the 3-dimensional part of the line element (\ref{quasi}), $dl^2 =
\gamma_{ij}dx^idx^j$,  in cartesian coordinates via the transformation
\begin{equation}\label{kugelq}
 (\tilde{r},\;\theta,\;\phi)=\left(\sqrt{x^2+y^2+z^2},\;\arccos\frac{z}{\tilde{r}},\;\arctan\frac{y}{x}\right),\quad \vct{x}_{sphere}=(\tilde{r},\;\theta,\;\phi)\,
\end{equation}
which results in 
\begin{equation}\label{cartmetric}
\dd
l^2=\left(\frac{\rho}{\tilde{r}}\right)^2\left[\frac{x^2+y^2\ee^{-2\mu}}{x^2+y^2}\dd
x^2+\frac{x^2\ee^{-2\mu}+y^2}{x^2+y^2}\dd y^2+\dd z^2+
\frac{2xy}{x^2+y^2}\left(1-\ee^{-2\mu}\right)\dd x\,\dd y\right]
\end{equation}
and make a series expansion of (\ref{cartmetric}) in powers of
$\tilde{r}=\sqrt{x^2+y^2+z^2}$ with the replacement of $\tilde{r}$ by
$r$ up to the orders $1/r^4$ and $a^2$. The reason for stopping at these
orders is the following: The expansion in $\frac{1}{r}$ corresponds to
an expansion in powers of $\frac{1}{c^2}$ with
$\frac{1}{r}\sim\frac{1}{c^2}$ and is clearly nothing else than a
post-Newtonian expansion. Notice that it is also possible to count in
terms of $m\sim\frac{1}{c^2}$ and $a\sim\frac{1}{c^2}$ instead of $1/r$
using $m=\frac{GM}{c^2}$ ($M$ is the mass in SI-units) and
$a=\frac{S}{Mc}$ where the well-known spin $S=Mca$ has been introduced
which for rapidly rotating black holes is of the order $1/c$. So an expansion up to the 4th order in $1/r$ will suffice to get some of the desired next-to-leading-order (self-) spin interaction Hamiltonians for a binary black hole system,
\begin{align}
\begin{aligned}
dl^2=&\left(\left(1+\frac{m}{2r}\right)^4+\frac{a^2}{2r^2}-\frac{ma^2}{2r^3}-\frac{m^2a^2}{8r^4}-\frac{a^2x^2}{r^4}+\frac{2a^2my^2}{r^5}-\frac{2a^2m^2y^2}{r^6}\right)\dd x^2\\
&+\left(\left(1+\frac{m}{2r}\right)^4+\frac{a^2}{2r^2}-\frac{ma^2}{2r^3}-\frac{m^2a^2}{8r^4}-\frac{a^2y^2}{r^4}+\frac{2a^2mx^2}{r^5}-\frac{2a^2m^2x^2}{r^6}\right)\dd y^2\\
&+\left(\left(1+\frac{m}{2r}\right)^4-\frac{a^2}{2r^2}+\frac{a^2z^2}{r^4}-\frac{ma^2}{2r^3}-\frac{m^2a^2}{8r^4}\right)\dd z^2\\
&+2\left(-\frac{xya^2}{r^4}-\frac{2a^2mxy}{r^5}+\frac{2a^2m^2xy}{r^6}\right)\,\dd x\,\dd y.
\end{aligned}
\end{align}

We will now put this line element into an invariant form with respect to
the direction of the spin. For that we define
\begin{equation}{\label{az}}
\vct{a}=\begin{pmatrix}
0\\
0\\
a
\end{pmatrix}\,, \quad \vct{a}  = (a^i) = (a_i)\; \quad
\Longrightarrow\quad\vct{e}_{z}\equiv \frac{\vct{a}}{a}\;,\;
z=\frac{\vct{a}\!\cdot\!\vct{x}}{a}=\frac{r\vct{a}\!\cdot\!\vct{n}}{a}\,,
\quad \vct{n}  = (n^i) = (n_i)
\end{equation}
and use the following relation
\begin{align}\label{e^2}
\begin{aligned}
x^2\dd y^2+y^2\dd x^2-2xy\,\dd x\,\dd y&=\left(\vct{e}_{z}\!\cdot\!(\vct{x}\times\dd\vct{x})\right)^2\\
&=\frac{1}{a^2}\epsilon_{ijk}\epsilon_{lpq}\,a^{i}x^{j}a^{l}x^{p}\,\dd x^{k}\dd x^{q}\\
&=\frac{1}{a^2}\bigg(a^2r^2\;\dd x^2+2(\vct{a}\!\cdot\!\vct{x})(\vct{a}\!\cdot\!\dd\vct{x})(\vct{x}\!\cdot\!\dd\vct{x})-(\vct{a}\!\cdot\!\vct{x})^2\dd x^2\\
&\qquad\quad-r^2(\vct{a}\!\cdot\!\dd\vct{x})^2-a^2(\vct{x}\!\cdot\!\dd\vct{x})^2\bigg)
\end{aligned}
\end{align}
with
\begin{equation}
\epsilon_{ijk}\epsilon_{lpq}=3!\,\delta^{[ijk]}_{[lpq]}=\delta^{i}_{l}\delta^{j}_{p}\delta^{k}_{q}+\delta^{i}_{q}\delta^{j}_{l}\delta^{k}_{p}+
\delta^{i}_{p}\delta^{j}_{q}\delta^{k}_{l}-\delta^{i}_{p}\delta^{j}_{l}\delta^{k}_{q}-\delta^{i}_{q}\delta^{j}_{p}\delta^{k}_{l}-
\delta^{i}_{l}\delta^{j}_{q}\delta^{k}_{p}\;,
\end{equation}
where $\epsilon_{ijk}$ is the completely antisymmetric Levi-Civita
tensor. Then we find 
\begin{equation}
\gamma_{ij}=\gamma_{ij}^{(s)}+\gamma_{ij}^{(2)}+\gamma_{ij}^{(3)}+\gamma_{ij}^{(4)}\,,
\end{equation}
where 
\begin{eqnarray}
\gamma_{ij}^{(s)}&=&\left(1+\frac{m}{2r}\right)^4\delta_{ij}\,, \\
\gamma_{ij}^{(2)}&=&\frac{1}{2}\frac{a^2}{r^2}\delta_{ij}-\frac{a^2n_{i}n_{j}}{r^2}-\frac{a_{i}a_{j}}{r^2}+
\frac{2(\vct{a}\!\cdot\!\vct{n})a_{(i}n_{j)}}{r^2}\,, \\
\gamma_{ij}^{(3)}&=&\left(\frac{3}{2}\frac{ma^2}{r^3}-
\frac{2m(\vct{a}\!\cdot\!\vct{n})^2}{r^3}\right)\delta_{ij}-\frac{2ma^2n_{i}n_{j}}{r^3}-\frac{2ma_{i}a_{j}}{r^3}+
\frac{4m(\vct{a}\!\cdot\!\vct{n})a_{(i}n_{j)}}{r^3}\,, \\
\gamma_{ij}^{(4)}&=&\left(-\frac{17}{8}\frac{m^2a^2}{r^4}+\frac{2m^2(\vct{a}\!\cdot\!\vct{n})^2}{r^4}\right)\delta_{ij}+
\frac{2m^2a^2n_{i}n_{j}}{r^4}+\frac{2m^2a_{i}a_{j}}{r^4}-\frac{4m^2(\vct{a}\!\cdot\!\vct{n})a_{(i}n_{j)}}{r^4}.
\end{eqnarray}
The index $(s)$ indicates the exact static limit.
The transformation of this metric to ADMTT coordinates up to the orders
$1/r^4$ and $a^2$ will be achieved by means of the transformation
formula (\ref{gijtrafo}) (the index qiso stands for quasi-isotropic)
\begin{equation}
\gamma_{ij}^{ADM}=\gamma_{ij}^{\text{qiso}}+\gamma_{ik}^{\text{qiso}}\,?\xi^{k}_{,\,j}?+\gamma_{jk}^{\text{qiso}}\,?\xi^{k}_{,\,i}?+
\gamma_{ij,\,k}^{\text{qiso}}\,\xi^{k}
\end{equation}
with vector
\begin{equation}\label{vecq}
\xi^{k}=-\frac{1}{4}\frac{a^2n^{k}}{r}+\frac{1}{2}\frac{(\vct{a}\!\cdot\!\vct{n})a^{k}}{r}-\frac{7}{16}\frac{m^2a^2n^{k}}{r^3}+
\frac{7}{4}\frac{m^2(\vct{a}\!\cdot\!\vct{n})^2n^{k}}{r^3}.
\end{equation}
This leads to
\begin{equation}
\gamma_{ij}^{ADM}=\left(\left(1+\frac{m}{2r}\right)^4+\frac{ma^2-3m(\vct{a}\!\cdot\!\vct{n})^2}{r^3}+
\frac{1}{2}\frac{m^2a^2}{r^4}-\frac{3m^2(\vct{a}\!\cdot\!\vct{n})^2}{r^4}\right)\delta_{ij}+h_{ij}^{TT}+\mathcal{O}\left(a^2,\frac{1}{r^5}\right)
\end{equation}
with
\begin{equation}\label{httq}
h_{ij}^{TT}=-\frac{7}{2}\frac{m^2a^2}{r^4}\delta_{ij}+7\frac{m^2(\vct{a}\!\cdot\!\vct{n})^2}{r^4}\delta_{ij}+
7\frac{m^2a^2n_{i}n_{j}}{r^4}-21\frac{m^2(\vct{a}\!\cdot\!\vct{n})^2n_{i}n_{j}}{r^4}+\frac{7}{2}\frac{m^2a_{i}a_{j}}{r^4}.
\end{equation}
The property of leading order $h_{ij}^{TT}$ of being quadratic in $a_i$
is nicely consistent with
\cite{garat}. In terms of a post-Newtonian expansion, we would call $h_{ij}^{TT}$
being of 4th post-Newtonian order, $1/c^8$.\\

\noindent
Now the calculations of the lapse and shift functions and the field momentum
density are going to be performed.
The part of the line element which contains the lapse and shift
functions reads (\ref{quasi}), 

\begin{align}\label{lsq}
\begin{aligned}
ds^2-\gamma_{ij}^{\text{qiso}}\dd x^{i}\dd
x^{j}&=\left(\left(\frac{\rho}{\tilde{r}}\right)^2\tilde{r}^2\sin^2\theta\,\ee^{-2\mu}(N^{\phi})^2-
N^2\right)\dd t^2+2\left(\frac{\rho}{\tilde{r}}\right)^2\tilde{r}^2\sin^2\theta\,\ee^{-2\mu}N^{\phi}\dd \phi\dd t\\
&=\left(N_{i}N^{i}-N^2\right)\dd t^2+2N_{i}\dd x^{i}\dd t\,.
\end{aligned}
\end{align}
Using 
\begin{equation}
\dd\phi=\frac{1}{\tilde{r}^2\sin^2\theta}\frac{\vct{a}}{a}\!\cdot\!\left(\vct{x}\times\dd\vct{x}\right)=
\frac{1}{a\tilde{r}^2\sin^2\theta}\epsilon_{ijk}a^{i}x^{j}\dd
x^{k} \label{dphi}\,, 
\end{equation}
the shift vector in quasi-isotropic coordinates is obtained in the form 
\begin{align}
\begin{aligned}
N_{i}^{\text{qiso}}&=\left(\frac{\rho}{\tilde{r}}\right)^2\frac{\ee^{-2\mu}}{a}N^{\phi}\epsilon^{ijk}a_{j}x_{k}\\
&=\left(-\frac{2m}{\tilde{r}^2}+\frac{2m}{\tilde{r}^3}-\frac{1}{2}\frac{(3m^3+ma^2)}{\tilde{r}^4}+
\frac{2m(\vct{a}\!\cdot\!\vct{n})^2}{\tilde{r}^4}\right)\epsilon_{ijk}a^{j}n^{k}
\end{aligned}
\end{align}
together with the lapse function 
\begin{align}
\begin{aligned}
N^{\text{qiso}}&=\sqrt{N_{i}^{\text{qiso}}N^{i}_{\text{qiso}}-g_{00}^{\text{qiso}}}\\
&=1-\frac{m}{\tilde{r}}+\frac{1}{2}\frac{m^2}{\tilde{r}^2}-\frac{1}{4}\frac{(m^3+ma^2)}{\tilde{r}^3}+
\frac{m(\vct{a}\!\cdot\!\vct{n})^2}{\tilde{r}^3}+\frac{1}{8}\frac{m^4}{\tilde{r}^4}+\frac{9}{4}\frac{m^2a^2}{\tilde{r}^4}-
\frac{4m^2(\vct{a}\!\cdot\!\vct{n})^2}{\tilde{r}^4}.
\end{aligned}
\end{align}
The ADM form of the shift function is obtained with the help of the
transformation with the vector $\xi^k$ taken from (\ref{vecq})
\begin{equation}
g_{0i}^{\text{ADM}}=g_{0i}^{\text{qiso}}+g_{k0}^{\text{qiso}}?\xi^{k}_{,\,i}?+g_{0i,\,k}^{\text{qiso}}\,\xi^{k}.
\end{equation}
The result reads, again substituting $r$ for $\tilde{r}$,
\begin{equation}\label{g0iqadm}
g_{0i}^{\text{ADM}}=N_{i}^{\text{ADM}}=\left(-\frac{2m}{r^2}+\frac{2m^2}{r^3}-\frac{3}{2}\frac{m^3}{r^4}+
\frac{5m(\vct{a}\!\cdot\!\vct{n})^2-ma^2}{r^4}\right)\epsilon_{ijk}a^{j}n^{k}
\end{equation}
or
\begin{equation}\label{shadm}
N^{i}_{\text{ADM}}=\left(-\frac{2m}{r^2}+\frac{6m^2}{r^3}-\frac{21}{2}\frac{m^3}{r^4}+
\frac{5m(\vct{a}\!\cdot\!\vct{n})^2-ma^2}{r^4}\right)\epsilon^{ijk}a_{j}n_{k},
\end{equation}
where use has been made of 
\begin{equation}\label{inversadm}
\gamma^{ij}_{ADM}=\left(1-\frac{2m}{r}+\frac{5}{2}\frac{m^2}{r^2}\right)\delta^{ij}+\mathcal{O}\left(\frac{1}{r^3}\right)\;.
\end{equation}

Next we determine the lapse function in ADM coordinates by means of the very easy formula (\ref{g00trafo}):

\begin{equation}
g_{00}^{\text{ADM}}=g_{00}^{\text{qiso}}+g_{00,\,k}^{\text{qiso}}\,\xi^{k}
\end{equation}
with
\begin{equation}
g_{00}^{\text{qiso}}=-1+\frac{2m}{r}-\frac{2m^2}{r^2}+\frac{3}{2}\frac{m^3}{r^3}+\frac{1}{2}\frac{ma^2}{r^3}-
\frac{2m(\vct{a}\!\cdot\!\vct{n})^2}{r^3}-\frac{m^4}{r^4}+\frac{6m^2(\vct{a}\!\cdot\!\vct{n})^2-m^2a^2}{r^4},
\end{equation}
resulting in
\begin{equation}\label{laadm}
N^{\text{ADM}}=1-\frac{m}{r}+\frac{1}{2}\frac{m^2}{r^2}-\frac{1}{4}\frac{m^3}{r^3}+\frac{1}{8}\frac{m^4}{r^4}+
\frac{1}{2}\frac{(3m(\vct{a}\!\cdot\!\vct{n})^2-ma^2)}{r^3}+\frac{1}{2}\frac{(5m^2a^2-9m^2(\vct{a}\!\cdot\!\vct{n})^2)}{r^4}\,.
\end{equation}
Hereof we calculate
\begin{equation}\label{wurzel}
\sqrt{-g}=N\sqrt{\gamma}=1+\frac{2m}{r}+\frac{5}{4}\frac{m^2}{r^2}+\mathcal{O}\left(\frac{1}{r^3}\right)
\end{equation}
and
\begin{equation}
g^{00}=-\frac{1}{N^2}=-1-\frac{2m}{r}-\frac{2m^2}{r^2}-\frac{3}{2}\frac{m^3}{r^3}-\frac{m^4}{r^4}+
\frac{3m(\vct{a}\!\cdot\!\vct{n})^2-ma^2}{r^3}+\frac{2m^2a^2}{r^4}\,.
\end{equation}

Finally, we calculate $\pi^{ij}$ via
\begin{equation}
\pi^{ij}=\sqrt{-g}\left(\gamma^{ip}\gamma^{jq}-\gamma^{ij}\gamma^{pq}\right)\Gamma_{pq}^{0}
\end{equation}
with
\begin{equation}\label{chris}
\Gamma_{pq}^{0}=\left(-\frac{6m}{r^3}-\frac{3}{2}\frac{m^3}{r^5}+\frac{35m(\vct{a}\!\cdot\!\vct{n})^2-
5ma^2}{r^5}\right)\epsilon_{kl(p}n_{q)}a^{k}n^{l}-\frac{10m(\vct{a}\!\cdot\!\vct{n})}{r^5}\epsilon_{kl(p}a_{q)}a^{k}n^{l}
\end{equation}
giving us the result

\begin{equation}\label{piquasi1}
\pi^{ij}=\left(-\frac{6m}{r^3}+\frac{12m^2}{r^4}-\frac{15m^3}{r^5}+\frac{35m(\vct{a}\!\cdot\!\vct{n})^2-
5ma^2}{r^5}\right)\epsilon^{kl(i}n^{j)}a_{k}n_{l}-\frac{10m(\vct{a}\!\cdot\!\vct{n})}{r^5}\epsilon^{kl(i}a^{j)}a_{k}n_{l}\,.
\end{equation}
We easily see that the gauge condition on $\pi^{ii}$ is already
satisfied in this order so there is no need to force $\xi^{0}$ to be unequal to zero.

\subsection{Transformation from harmonic to ADMTT coordinates}

First we need the Kerr-metric in harmonic coordinates at least up to the
desired order. To achieve this goal let us start with Boyer-Lindquist coordinates \cite{boyer}

\begin{equation}
\dd s^2=-\frac{\Delta}{\rho^{2}}\left[\dd t-a\sin^{2}\theta
\,\dd\phi\right]^{2}+
\frac{\sin^2\theta}{\rho^{2}}\left[(r^2+a^2)\dd\phi-a\,\dd t\right]^2+\frac{\rho^2}{\Delta}\dd r^2+\rho^2\,\dd\theta^2
\end{equation} 
with
\begin{center}
 $\Delta\equiv r^2-2mr+a^2$ \qquad and\qquad $\rho^{2}\equiv r^2+a^2\cos^{2}\theta$\,.
\end{center}
Rewritten, the line element reads
\begin{align}
\begin{aligned}
\dd s^2&=\frac{a^2\sin^2\theta_{B}-\Delta}{\rho^2}\dd t_{B}^2+
\frac{2\sin^2\theta_{B}}{\rho^2}\left(\Delta a-a(r_{B}^2+a^2)\right)\dd\phi_{B}\,\dd t_{B}\\
&\quad+\frac{\rho^2}{\Delta}\dd
r_{B}^2+\rho^2\dd\theta_{B}^2+\frac{\sin^2\theta_{B}}{\rho^2}\left[(r_{B}^2+a^2)-
\Delta a^2\sin^2\theta_{B}\right]\dd\phi_{B}^2
\end{aligned}
\end{align}
with $(t_{B},\,r_{B},\,\theta_{B},\,\phi_{B})$ labeling the
Boyer-Lindquist coordinates .\\

The harmonic coordinates condition $\Box
x^{\nu}_{H}=D_{\mu}D^{\mu}x^{\nu}_{H}=0$ (label H for harmonic) gives
rise to the equations \cite{cookharmonic, cookinitial}:
\begin{equation}
t_{H}=t_{B}\,,
 \end{equation}
and 
\begin{eqnarray}\label{phiharm}
x_{H}+iy_{H}&=&(r_{B}-m+ia)e^{i\phi}\sin{\theta_{B}}\\
\phi&=&\phi_{B}+\frac{a}{r_{+}-r_{-}}\ln\left|\frac{r_{B}-r_{+}}{r_{B}-r_{-}}\right|\qquad\text{(be
aware $\phi\neq{\phi_{H}}$)}, 
\end{eqnarray}
where
\begin{eqnarray}
r_{\pm}&=&m\pm\sqrt{m^2-a^2}\\
r^{2}_{H}&=&x^{2}_{H}+y^{2}_{H}+z^{2}_{H}=(r_{B}-m)^2+a^2\sin^{2}\theta_{B}\qquad \text{with}\\
z_{H}^{2}&=&r^{2}_{H}\cos^{2}_{H}=(r_{B}-m)^2\cos^{2}\theta_{B}\,.
\end{eqnarray}
These equations are solved perturbatively so that we get the
Boyer-Lindquist coordinates depending on the harmonic ones,

\begin{align}
 r_{B}&=r_{H}+m-\frac{a^2\sin^2\theta_{H}}{2r_{H}}+\frac{a^4\left(3+5\cos(2\theta_{H})\right)\sin^2\theta_{H}}{16r_{H}^3}+
 O\left(a^6,\frac{1}{r_{H}^5}\right)\,,\\
 \theta_{B}&=\theta_{H}-\frac{a^2\sin\theta_{H}\cos\theta_{H}}{2r_{H}^2}+
 \frac{3a^4\sin\theta_{H}\cos\theta_{H}\cos(2\theta_{H})}{8r_{H}^4}+O\left(a^6,\frac{1}{r_{H}^6}\right)\,,\\
 \phi_{B}&=\phi_{H}+\frac{am^2}{3r_{H}^3}+O\left(a^5,\frac{1}{r_{H}^5}\right)\,.
\end{align}
The line element in harmonic coordinates then reads
\begin{equation}\label{linieharm}
\dd s^2=g_{00}\,\dd t^2+2g_{0\phi}\dd t\,\dd\phi+\dd l^2_{a}+\dd
l^2_{a^2}\,,\textrm{\quad from here on: \quad} (r_{H},\theta_{H},\phi_{H})\equiv(r,\theta,\phi)
\end{equation}
with the expressions
\begin{equation}
g_{00}=-1+\frac{2m}{r}-\frac{2m^2}{r^2}-\frac{ma^2-4m^3+3a^2m\cos(2\theta)}{2r^3}+\frac{2m^2a^2-2m^4+4m^2a^2\cos(2\theta)}{r^4}\,,
\end{equation}
\begin{equation}
g_{0\phi}=-\frac{2am\sin^2\theta}{r}+\frac{2am^2\sin^2\theta}{r^2}+\frac{am\sin^2\theta\,(3a^2-4m^2+5a^2\cos(2\theta))}{2r^3}-
\frac{am^2\sin^2\theta\,(3a^2-2m^2+5a^2\cos(2\theta))}{r^4}\,,
\end{equation}
\begin{equation}
\dd
l^2_{a}=\left(1+\frac{2m}{r}+\frac{2m^2}{r^2}+\frac{2m^3}{r^3}+\frac{2m^4}{r^4}\right)\dd
r^2+(r+m)^2\dd\theta^2+
(r+m)^2\sin^2{\theta}\,\dd\phi^2-\frac{2am^2\sin^2{\theta}}{r^2}\left(1+\frac{2m}{r}\right)\dd r\,\dd\phi\,,
\end{equation}

\begin{equation}
\begin{split}
\dd
l^2_{a^2}=&\left(\frac{a^2m}{r^3}(3\sin^2\theta-2)-\frac{a^2m^2}{2r^4}(1+3\cos(2\theta))\right)\dd
r^2-
\frac{a^2m}{r}\left(\frac{1+3\cos(2\theta)}{2}+\frac{m\cos(2\theta)}{r}\right)\dd\theta^2\\
&+\frac{a^2m}{2r}\sin^2\theta\,\left(-(1+3\cos(2\theta))+\frac{m(\cos(2\theta)-3)}{r}\right)\dd\phi^2-
\frac{2m^2a^2\sin(2\theta)}{2r^3}\dd r\,\dd \theta\,.
\end{split}
\end{equation}
$\dd l^2_{a}$ and $\dd l^2_{a^2}$ are the pure spatial parts of the line element, linear and quadratic in $a$ respectively.
Now the harmonic 3-metric
$\gamma_{ij}=\gamma_{ij}^{(a)}+\gamma_{ij}^{(a^2)}$ related respectively
to $\dd l^2_{a}$ and $\dd l^2_{a^2}$ will be calculated.

The transformation formula of the metric tensor gives for $\gamma_{ij}^{(a)}$ the result
\begin{equation}
\gamma_{ij}^{(a)}=\left(1+\frac{2m}{r}+\frac{m^2}{r^2}\right)\delta_{ij}+\left(\frac{m^2}{r^2}+\frac{2m^3}{r^3}+
\frac{2m^4}{r^4}\right)n_{i}n_{j}-\left(\frac{2m^2}{r^3}+\frac{4m^3}{r^4}\right)\epsilon_{kl(i}n_{j)}a^{k}n^{l}\,, \label{gammaa}
\end{equation}
and for $\gamma_{ij}^{(a^2)}$:
\begin{align}
\begin{aligned}
\gamma_{ij}^{(a^2)}&=\left(\frac{ma^2-3m\left(\vct{a}\!\cdot\!\vct{n}\right)^2}{r^3}+\frac{m^2\left(\vct{a}\!\cdot\!\vct{n}\right)^2-
2m^2a^2}{r^4}\right)\delta_{ij}-\frac{4m^2\left(\vct{a}\!\cdot\!\vct{n}\right)a_{(i}n_{j)}}{r^4}+\frac{3m^2a_{i}a_{j}}{r^4}\\
&\quad+\frac{3m^2a^2n_{i}n_{j}}{r^4}-\frac{3m^2\left(\vct{a}\!\cdot\!\vct{n}\right)^2n_{i}n_{j}}{r^4}\,.
\end{aligned}
\end{align}

From the $g_{0\phi}$ part we get the shift vector in the harmonic gauge
\begin{equation}
N_{i}^{H}=\left(-\frac{2m}{r^2}+\frac{2m^2}{r^3}-\frac{2m^3}{r^4}+\frac{5m\left(\vct{a}\!\cdot\!\vct{n}\right)^2-ma^2}{r^4}\right)\epsilon_{ijk}a^{j}n^{k}\,.
\end{equation}
The lapse function in harmonic coordinates turns out to be
\begin{equation}
N^{H}=1-\frac{m}{r}+\frac{m^2}{2r^2}-\frac{m^3}{2r^3}+\frac{3m^4}{8r^4}+\frac{-a^2m+3m\left(\vct{a}\!\cdot\!\vct{n}\right)^2}{2r^3}+
\frac{5a^2m^2-9m^2\left(\vct{a}\!\cdot\!\vct{n}\right)^2}{2r^4}\,.
\end{equation}

The 3-metric will be brought into ADM-form by means of the transformation formula (\ref{gijtrafo}):
\begin{equation}
\gamma_{ij}^{\text{ADM}}=\gamma_{ij}+\gamma_{ij,\,k}\;\xi^{k}+\gamma_{ki}\;?\xi^{k}_{,\,j}?+\gamma_{kj}\;?\xi^{k}_{,\,j}?+\gamma_{kl}\;?\xi^{k}_{,\,i}?\,?\xi^{l}_{,\,j}?
\end{equation}
with vector
\begin{equation}
\xi^{i}=\frac{1}{4}\frac{m^2}{r}n^{i}-\frac{1}{3}\frac{m^2}{r^2}\epsilon^{kli}a_{k}n_{l}-\frac{1}{2}\frac{m^2}{r^3}a^2n^{i}+\frac{1}{4}\frac{m^2(\vct{a}\!\cdot\!\vct{n})a^{i}}{r^3}+\frac{3}{2}\frac{m^2}{r^3}(\vct{a}\!\cdot\!\vct{n})^2n^{i}\,.
\end{equation}

The transformation of $g_{00}^{\text{H}}$ is provided by the formula (\ref{g00trafo}):
\begin{equation}
g_{00}^{\text{ADM}}=g_{00}^{\text{H}}+g_{00,\,i}^{\text{H}}\xi^{i}
\end{equation}
with
\begin{equation}
g_{00}^{\text{H}}=-1+\frac{2m}{r}-\frac{2m^2}{r^2}+\frac{2m^3}{r^3}-\frac{2m^4}{r^4}+\frac{ma^2-3m(\vct{a}\!\cdot\!\vct{n})^2}{r^3}+\frac{8m^2(\vct{a}\!\cdot\!\vct{n})^2-2m^2a^2}{r^4}\,.
\end{equation}
This results in
\begin{equation}
g_{00}^{\text{ADM}}=-1+\frac{2m}{r}-\frac{2m^2}{r^2}+\frac{3}{2}\frac{m^3}{r^3}+\frac{ma^2-3m(\vct{a}\!\cdot\!\vct{n})^2}{r^3}-\frac{m^4}{r^4}+\frac{8m^2(\vct{a}\!\cdot\!\vct{n})^2-2m^2a^2}{r^4}\,.
\end{equation}
The transformation of $g_{0i}^{\text{ADM}}$ is provided by the formula (\ref{g0itrafo}):
\begin{equation}
g_{0i}^{\text{ADM}}=g_{0i}^{\text{H}}+g_{j0}^{\text{H}}\,?\xi^{j}_{,\,i}?+g_{0i,\,j}^{\text{H}}\,\xi^{j}
\end{equation}
which results in
\begin{equation}
g_{0i}^{\text{ADM}}=\left(-\frac{2m}{r^2}+\frac{2m^2}{r^3}-\frac{3}{2}\frac{m^3}{r^4}+\frac{5m(\vct{a}\!\cdot\!\vct{n})^2-ma^2}{r^4}\right)\epsilon_{ijk}a^{j}n^{k}
\end{equation}
giving the same $\pi^{ij}$ as above.

\section{Source terms for Kerr metric in ADMTT coordinates}
\label{sec3}

The source of the Hamiltonian constraint for a single Kerr black hole
presented perturbatively in ADMTT coordinates is given by
\begin{equation}\label{phisource}
-\Delta\phi=\left[1-\frac{1}{8}\phi+\frac{1}{64}\phi^{2}-\frac{1}{512}\phi^{3}\right]
16\pi m\delta(\vct{x})
+\left(-\frac{1}{2}+\frac{1}{16}\phi\right)\widehat{\left(\vct{a}\!\cdot\!\vct{\partial}\right)^2}16\pi
m \delta(\vct{x})+\left(\pi^{ij}\right)^2+\mathcal{O}\left(\frac{1}{r^7}\right)
\end{equation}
with
$\widehat{\left(\vct{a}\!\cdot\!\vct{\partial}\right)^2}:=\widehat{a^{i}a^{j}}\partial_{i}\partial_{j}=\left(\vct{a}\!\cdot\!\vct{\partial}\right)^2-\frac{1}{3}a^2\Delta$
being tracefree to get rid of the delta-function which appears by
calculating $\partial_{i}\partial_{j}\frac{1}{r}$ and end up with the quadrupole moment of a Kerr black hole (with the specific factor $1/2$), which describes its deformation. This source is
sufficient to reproduce the isotropic part of $\gamma_{ij}$. Notice that
$\left(\pi^{ij}\right)^2$ is given by the Bowen-York expression \cite{bowen} and that $\phi^{n}\delta(\vct{x})$ with $n\in\mathbbm{N}\setminus\{0\}$ is regularized to zero for a single black hole but will contribute for a binary system. From Eq. (95) we readily read off,
\begin{equation}
E^{*}=m\left(1-\frac{1}{2}\widehat{\left(\vct{a}\!\cdot\!\vct{\partial}\right)^2}\right)\delta(\vct{x})\,.
\end{equation}

The source for momentum constraint can be calculated to give
\begin{equation}\label{pisource}
\pi^{ij}_{,\,j}=\frac{1}{4}\epsilon^{ikl}a_{k}\partial_{l}\left(1-\frac{1}{6}\widehat{(\vct{a}\!\cdot\!\vct{\partial})^2}\right)
16\pi m\delta(\vct{x})+\left(-\frac{1}{2}+\frac{1}{16}\phi\right)\phi_{,\,j}\pi^{ij}+\mathcal{O}\left(\frac{1}{r^7}\right)
\end{equation}
which can be checked by using the formula
\begin{equation}
\pi^{ij}=\Theta^{ij}_{k}A^{k}\qquad \pi^{ij}_{,\,j}=A^{i}
\end{equation}
with
\begin{equation}
\Theta^{ij}_{k}:=\left(-\frac{1}{2}\delta_{ij}\partial_{k}+\delta_{ik}\partial_{j}+\delta_{jk}\partial_{i}-\frac{1}{2}\partial_{i}\partial_{j}\partial_{k}\Delta^{-1}\right)\Delta^{-1}\,.
\end{equation}
From the Eq. (97) we find the source expression

\begin{equation}
S^{*i}=-\frac{m}{2}\epsilon^{kli}a_{k}\partial_{l}\left(1-\frac{1}{6}\widehat{\left(\vct{a}\!\cdot\!\vct{\partial}\right)^2}\right)\delta(\vct{x})\,.
\end{equation}

The source for $h_{ij}^{TT}$ in the leading order comes by the 
Ricci equation $R_{ij}=0$ which reads, to leading order, 

\begin{align}
\begin{aligned}
\Delta
h_{ij}^{TT} =\left[\frac{1}{8}\phi_{,\,i}\,\phi_{,\,j} +N^{k}\left(N_{i,\,jk}+N_{j,\,ik}-N_{k,\,ij}\right)
-\frac{1}{2}\Delta(N_{i}N_{j}) -2\frac{N_{,\,ij}}{N}
+ \frac{1}{2}\left(\phi_{,\,j}\,N_{,\,i}+\phi_{,\,i}\,N_{,\,j}\right)\right]^{TT}
\end{aligned}
\end{align}
with
\begin{align}
\begin{aligned}
\phi&=-\Delta^{-1}\left[16\pi m\left(1-\frac{1}{2}\widehat{\left(\vct{a}\!\cdot\!\vct{\partial}\right)^2}\right)\,\delta(\vct{x})\right]\\
&=\frac{4m}{r}+\frac{2ma^2-6m(\vct{a}\!\cdot\!\vct{n})^2}{r^3}
\end{aligned}
\end{align}
and 
\begin{align}
\begin{aligned}
N= 1-\frac{1}{4} \phi + \frac{1}{32} \phi^2 - \frac{1}{256} \phi^3 + \frac{1}{2048} \phi^4 + \Delta^{-1} (\pi^{ij})^2 + \mathcal{O}\left(\frac{1}{r^5}\right)
\end{aligned}
\end{align}
as well as  
\begin{align}
\begin{aligned}
N_i = N^{i} = 16\pi\Delta^{-1} S^{*i} = 16\pi\Delta^{-1}\left(-\frac{1}{2}m\epsilon^{kli}a_{k}\partial_{l}\delta(\vct{x})\right)\,.
\end{aligned}
\end{align}
Notice that our $h_{ij}^{TT}$ is also the result of the equation
\begin{equation}
\Delta h_{ij}^{TT}=\delta_{ij}^{TTkl}\left(\frac{7(3m^2a_{k}a_{l}-m^2a^2\delta_{kl})}{r^6}\right)=\delta_{ij}^{TTkl}\left(\frac{21m^2a_{k}a_{l}}{r^6}\right)\,,
\end{equation}
i.e. $a_{k}a_{l}/r^6$ is an irreducible part of a metric of the form
(\ref{mansatz}) to be TT-projected. This result implies that the
regularized source term in Eq. (26), reading
$E^*({S^*}_k/E^*) ({S^*}_l/E^*)$, vanishes, i.e.
$[(S^*_k/E^*)(S^*_l/E^*)]_{reg} =  [(S^*_k/E^*)]_{reg}[(S^*_l/E^*)]_{reg}
=0$. Technically one may argue that in the regularization procedure one
is allowed to apply the ``tweedling of products'' by Infeld and
Pleba\'nski, \cite{IP}, and to put to zero each term seperately. Such a
regularization property is particularly valid in dimensional
regularization, \cite{DJS01}. On the other side, an expansion like 
Eq. (26) would never be expected to represent highly nonlinear terms
correctly. So the vanishing of the expression in question is quite satisfying.
The outcome of our considerations is the vanishing of $S^{*ij}$ in the
given approximation, i.e.    
 
\begin{equation}
S^{*ij}=0\,.
\end{equation}
\subsection{Leading-order sources for the constraints for boosted Kerr}

To obtain leading order source terms with non-vanishing linear momentum
we substitute in the constraint equations for point particles
\cite{JS98} $\vct{p}\delta(\vct{x})$ by
$(\vct{p}-\frac{1}{2}m\vct{a}\times\vct{\nabla})\delta(\vct{x})$
according to the a remark in \cite{schafermagnetic} (in the following equation point-particle expressions are denoted
by the upper index $(PP)$):

\begin{equation}
 \begin{split}
 -\Delta\phi =-\Delta\phi_{\vct{p}=0}^{(PP)}+\left(-\frac{1}{2}+\frac{5}{16}\phi^{(PP)}_{\vct{p}=0}\right)\vct{p}\!\cdot\!\left(\vct{a}\times\vct{\nabla}\right)16\pi\delta(\vct{x}),
 \end{split}
 \end{equation}

\begin{equation}
\pi^{ij}_{,\,j}= - {8}\pi p^{i}\delta(\vct{x})+4\pi m\left(\vct{a}\times\vct{\nabla}\right)^{i}\delta(\vct{x})
\end{equation}
[${\bf p} = (p_i)$ and
$\left(\vct{a}\times\vct{\nabla}\right)^{i}=\epsilon^{ijk}a_j\partial_k
= \left(\vct{a}\times\vct{\nabla}\right)_{i}$].
Note that the term in Eq. (107) containing the factor $5/16$ is already needed
at leading order because the integral for the Hamiltonian of the term
with factor $1/2$ vanishes; however, the latter term contributes
by iteration via linear momentum independent expressions.    

For a boosted Kerr metric in ADMTT coordinates we can then state the
following source expression, to linear order in the momentum $\vct{p}$
and to 6th order in $1/r$ of the expansion of the
unboosted Kerr metric, see Eqs. (95) and (97), 

\begin{equation}
\begin{split}
-\Delta\phi=&\;\left[1-\frac{1}{8}\phi+\frac{1}{64}\phi^{2}-\frac{1}{512}\phi^{3}\right]16\pi
m \delta(\vct{x})
+\left(-\frac{1}{2}+\frac{1}{16}\phi\right)\widehat{\left(\vct{a}\!\cdot\!\vct{\partial}\right)^2}16\pi
m\delta(\vct{x}) +\left(\pi^{ij}\right)^2   \\
&\;+\left(-\frac{1}{2}+\frac{5}{16}\phi\right)\vct{p}\!\cdot\!\left(\vct{a}\times\vct{\nabla}\right)16\pi\delta(\vct{x}),
\end{split}
\end{equation}

\begin{equation}
\begin{split}
\pi^{ij}_{,\,j}=&\frac{1}{4}\epsilon^{ikl}a_{k}\partial_{l}
\left(1-\frac{1}{6}\widehat{(\vct{a}\!\cdot\!\vct{\partial})^2}\right){16}\pi
m\delta(\vct{x}) +\left(-\frac{1}{2}+\frac{1}{16}\phi\right)\phi_{,\,j}\pi^{ij} - {8}\pi p^{i}\delta(\vct{x}).
\end{split}
\end{equation}
Here we had to take into account that the substitution in the momentum
constraint leads to no new information regarding the spin because the
expansion of the Kerr metric already gives us the correct linear-in-spin term
which has no direct connection with the linear momentum. This can be
regarded as consistency check for our specific substitution.

\subsection{Source terms in the CFC reduced Kerr metric}

It is straightforward to calculate the source terms in the Eqs. (20)-(22) from our known form of the Kerr metric. The outcome reads

\begin{equation}
E^{*}=m\left(1-\frac{1}{2}\widehat{\left(\vct{a}\!\cdot\!\vct{\partial}\right)^2}\right)\delta(\vct{x})\,,
\end{equation}

\begin{equation}\label{sstern}
S^{*}=0\,,
\end{equation}
and 

\begin{equation}
S^{*i}=-\frac{m}{2}\epsilon^{kli}a_{k}\partial_{l}\left(1-\frac{1}{6}\widehat{\left(\vct{a}\!\cdot\!\vct{\partial}\right)^2}\right)\delta(\vct{x})\,.
\end{equation}
All source terms are in full agreement with our previous findings. 

\section{Spin interaction Hamiltonians}
\label{sec4}

The Hamiltonian and momentum constraint equations for point masses are
given in the form, see, e.g. \cite{JS98}

\begin{eqnarray}
\gamma^{-1/2}\left[\gamma
R+\frac{1}{2}\left(g_{ij}\pi^{ij}\right)-\pi_{ij}\pi^{ij}\right]&=&16\pi
\sum_{a}\left(\gamma^{ij}p_{ai}p_{aj}+m_{a}^2\right)^{1/2}\delta_{a}\label{superh}\,,\\
-?\pi^{ij}_{;\,j}?&=&8\pi\sum_{a}\gamma^{ij}p_{aj}\delta_{a}. \label{superp}
\end{eqnarray}
The vectors $\vct{x}_{a}=(x^i_{a})\in \mathbbm{R}$ denote the position
of the $a$-th point mass, also we define
$\vct{r}_{a}:=\vct{x}-\vct{x}_{a}$, $r_{a}:=|\vct{r}_{a}|$,
$\vct{n}_{a}:=\vct{r}_{a}/r_{a}$, and for $a\neq b$,
$\vct{r}_{ab}:=\vct{x}_{a}-\vct{x}_{b}$. The linear
momentum vector is denoted by $\vct{p}_{a}=(p_{ai})$ and the short-cut
$\delta_{a}$ has the meaning $\delta (\vct{x}-\vct{x}_{a})$. The
ADM Hamiltonian takes the form
\begin{equation}\label{hamformel}
H\left[\vct{x}_{a},\vct{p}_{a},h_{ij}^{TT},\pi^{ij}_{TT}\right]=-
\frac{1}{16\pi}\int\dd^{3}x\;
\Delta\phi\left[\vct{x}_{a},\vct{p}_{a},h_{ij}^{TT},\pi^{ij}_{TT}\right]\,.
\end{equation}

\subsection{Linear in $G$ spin-spin interaction Hamiltonians}

Our source for the Hamiltonian constraint is generalized to two black holes by superposition

\begin{equation}
-\Delta\phi_{\text{I}}=16\pi\left(1-\frac{1}{2}\widehat{(\vct{a}_{1}\!\cdot\!\vct{\partial}_{1})^2}\right)m_{1}\delta_{1}+
16\pi\left(1-\frac{1}{2}\widehat{(\vct{a}_{2}\!\cdot\!\vct{\partial}_{2})^2}\right)m_{2}\delta_{2}\;.
\end{equation}

From this expression the Hamiltonian results in the form

\begin{equation}
H_{\text{I}}=-\frac{1}{16\pi}\int\Delta\phi_{\text{I}}\,\dd^{3} x=m_{1}+m_{2}
\end{equation}

being the rest-mass energy.

For next order Hamiltonian we must calculate $\phi_{\text{I}}$

\begin{equation}\label{phi1}
\phi_{\text{I}}=4\left(1-\frac{1}{2}\widehat{(\vct{a}_{1}\!\cdot\!\vct{\partial}_{1})^2}\right)\frac{m_{1}}{r_{1}}+
4\left(1-\frac{1}{2}\widehat{(\vct{a}_{2}\!\cdot\!\vct{\partial}_{2})^2}\right)\frac{m_{2}}{r_{2}}\,.
\end{equation}

The next order source takes then the form

\begin{equation}
\begin{split}
-\Delta\phi_{\text{II}_{1}}=&-\frac{1}{8}\phi_{\text{I}}\left[16\pi\left(1-\frac{1}{2}\widehat{(\vct{a}_{1}\!\cdot\!\vct{\partial}_{1})^2}\right)m_{1}\delta_{1}
+16\pi\left(1-\frac{1}{2}\widehat{(\vct{a}_{1}\!\cdot\!\vct{\partial}_{2})^2}\right)m_{2}\delta_{2}\right]\\
=&16\pi\left(-\frac{1}{2}\frac{m_{1}}{r_{1}}+\frac{m_{1}}{4}\widehat{(\vct{a}_{1}\!\cdot\!\vct{\partial}_{1})^2}\frac{1}{r_{1}}\right)m_{2}
\left(1-\frac{1}{2}\widehat{(\vct{a}_{2}\!\cdot\!\vct{\partial}_{2})^2}\right)\delta_{2}\\
&+16\pi\left(-\frac{1}{2}\frac{m_{2}}{r_{2}}+\frac{m_{2}}{4}\widehat{(\vct{a}_{2}\!\cdot\!\vct{\partial}_{2})^2}\frac{1}{r_{2}}\right)m_{1}
\left(1-\frac{1}{2}\widehat{(\vct{a}_{1}\!\cdot\!\vct{\partial}_{1})^2}\right)\delta_{1}\,,
\end{split}
\end{equation}

after Hadamard regularization which disposes of the self-terms like $\frac{1}{r_{1}}\delta_{1}$ and $\frac{1}{r_{1}}\partial_{i}\partial_{j}\delta_{1}$.

Integration then gives

\begin{align}
\begin{aligned}
H_{\text{II}_{1}}=-\frac{1}{16\pi}\int\Delta\phi_{\text{II}_{1}}\,\dd^{3}x&=-\frac{m_{1}m_{2}}{r_{12}}+
\frac{1}{2}m_{1}m_{2}\widehat{(\vct{a}_{2}\!\cdot\!\vct{\partial}_{2})^2}\frac{1}{r_{12}}+
\frac{1}{2}m_{1}m_{2}\widehat{(\vct{a}_{1}\!\cdot\!\vct{\partial}_{1})^2}\frac{1}{r_{12}}\\
&\quad-\frac{1}{4}m_{1}m_{2}\widehat{(\vct{a}_{1}\!\cdot\!\vct{\partial}_{1})^2}\widehat{(\vct{a}_{2}\!\cdot\!\vct{\partial}_{2})^2}\frac{1}{r_{12}}\\
&=H_{N} + H_{S_{1}^{2}} +H_{S_{2}^{2}} + H_{S_{1}^2S_{2}^2}
\end{aligned}
\end{align}
with
\begin{eqnarray}
H_{N}&=&-\frac{m_{1}m_{2}}{r_{12}},\\
H_{S_{1}^{2}} + H_{S_{2}^{2}} &=&\frac{1}{2}\frac{m_{1}m_{2}}{r_{12}^3}\left(3\left(\vct{a}_{1}\!\cdot\!\vct{n}_{12}\right)^2+
3\left(\vct{a}_{2}\!\cdot\!\vct{n}_{12}\right)^2-\vct{a}_{1}^2-\vct{a}_{2}^2\right),\\
H_{S_{1}^2S_{2}^2}&=&-\frac{3}{2}\frac{m_{1}m_{2}}{r_{12}^5}\left(\left(\vct{a}_{1}\!\cdot\!\vct{a}_{2}\right)^2+\frac{1}{2}
\vct{a}_{1}^2\vct{a}_{2}^2\right)+\frac{15}{4}\frac{m_{1}m_{2}}{r_{12}^5}\left(\vct{a}_{1}^2\left(\vct{a}_{2}\!\cdot\!\vct{n}_{12}\right)^2+
\vct{a}_{2}^2\left(\vct{a}_{1}\!\cdot\!\vct{n}_{12}\right)^2\right)\\
&
&+\frac{15m_{1}m_{2}\left(\vct{a}_{1}\!\cdot\!\vct{n}_{12}\right)\left(\vct{a}_{2}\!\cdot\!\vct{n}_{12}\right)}{r_{12}^5}
\left(\vct{a}_{1}\!\cdot\!\vct{a}_{2}-\frac{7}{4}\left(\vct{a}_{1}\!\cdot\!\vct{n}_{12}\right)\left(\vct{a}_{2}\!\cdot\!\vct{n}_{12}\right)\right)\nonumber\,.
\end{eqnarray}

Now we come to the contribution of $(\pi^{ij})^2$ leading to some new
Hamiltonians. The momentum of the binary system
in the leading order is given by linear superposition

\begin{equation}
(\pi^{ij})^2=(\pi^{ij}_{1})^2+(\pi^{ij}_{2})^2+2\pi^{ij}_{1}\pi^{ij}_{2}
\end{equation}

with

\begin{align}
\pi^{ij}_{1}&=-m_{1}\epsilon^{ipl}\partial^{j}_{1}a_{1p}\partial_{1l}\left(1-\frac{1}{6}\widehat{(\vct{a}_{1}\!\cdot\!\vct{\partial}_{1})^2}\right)
\frac{1}{r_{1}}-m_{1}\epsilon^{jpl}\partial^{i}_{1}a_{1p}\partial_{1l}\left(1-\frac{1}{6}\widehat{(\vct{a}_{1}\!\cdot\!\vct{\partial}_{1})^2}\right)
\frac{1}{r_{1}}\;,\\
\pi^{ij}_{2}&=\pi^{ij}_{1}(1\leftrightarrow 2)\,.
\end{align}

The interaction terms will only appear in the object

\begin{align}
\begin{aligned}
2\pi^{ij}_{1}\pi^{ij}_{2}&=4m_{1}m_{2}\epsilon^{ipl}a_{1p}\partial_{1l}\partial_{1j}
\left(1-\frac{1}{6}\widehat{(\vct{a}_{1}\!\cdot\!\vct{\partial}_{1})^2}\right)\frac{1}{r_{1}}\epsilon^{iqs}a_{2q}
\partial_{2s}\partial_{2j}\left(1-\frac{1}{6}\widehat{(\vct{a}_{2}\!\cdot\!\vct{\partial}_{2})^2}\right)\frac{1}{r_{2}}\\
&\quad+4m_{1}m_{2}\epsilon^{ipl}a_{1p}\partial_{1l}\partial_{1j}\left(1-\frac{1}{6}\widehat{(\vct{a}_{1}\!\cdot\!\vct{\partial}_{1})^2}\right)
\frac{1}{r_{1}}\epsilon^{jqs}a_{2q}\partial_{2s}\partial_{2i}\left(1-\frac{1}{6}\widehat{(\vct{a}_{2}\!\cdot\!\vct{\partial}_{2})^2}\right)\frac{1}{r_{2}}
\end{aligned}
\end{align}

leading to a Hamiltonian

\begin{align}
\begin{aligned}
H_{\text{II}_{2}}&=-\frac{1}{16\pi}\int\Delta\phi_{\text{II}_{2}}\,\dd^{3} x\\
&=-\frac{1}{2}m_{1}m_{2}\epsilon^{ipl}a_{1p}\partial_{1l}\partial_{1j}\left(1-\frac{1}{6}\widehat{(\vct{a}_{1}\!\cdot\!\vct{\partial}_{1})^2}\right)
\epsilon^{iqs}a_{2q}\partial_{2s}\partial_{2j}\left(1-\frac{1}{6}\widehat{(\vct{a}_{2}\!\cdot\!\vct{\partial}_{2})^2}\right)r_{12}\\
&\quad-\frac{1}{2}m_{1}m_{2}\epsilon^{ipl}a_{1p}\partial_{1l}\partial_{1j}\left(1-\frac{1}{6}\widehat{(\vct{a}_{1}\!\cdot\!\vct{\partial}_{1})^2}\right)
\epsilon^{jqs}a_{2q}\partial_{2s}\partial_{2i}\left(1-\frac{1}{6}\widehat{(\vct{a}_{2}\!\cdot\!\vct{\partial}_{2})^2}\right)r_{12}\\\\
&=-\frac{1}{2}m_{1}m_{2}\epsilon^{ipl}a_{1p}\partial_{1l}\partial_{1j}\left(1-\frac{1}{6}\widehat{(\vct{a}_{1}\!\cdot\!\vct{\partial}_{1})^2}\right)
\epsilon^{iqs}a_{2q}\partial_{2s}\partial_{2j}\left(1-\frac{1}{6}\widehat{(\vct{a}_{2}\!\cdot\!\vct{\partial}_{2})^2}\right)r_{12}\\\\
&=H_{S_{1}S_{2}}+H_{S_{1}S_{2}^3}+H_{S_{2}S_{1}^3}
\end{aligned}
\end{align}
(notice  $\partial_{1i}=-\partial_{2i}$).

We then get

\begin{align}
\begin{aligned}
H_{S_{1}S_{2}}&=-\frac{1}{2}m_{1}m_{2}\left(\vct{a}_{1}\!\cdot\!\vct{a}_{2}(\vct{\partial}_{1}\!\cdot\!\vct{\partial}_{2})^2-
(\vct{a}_{1}\!\cdot\!\vct{\partial}_{2})(\vct{a}_{2}\!\cdot\!\vct{\partial}_{1})(\vct{\partial}_{1}\!\cdot\!\vct{\partial}_{2})\right)r_{12}\\
&=-m_{1}m_{2}\frac{\vct{a}_{1}\!\cdot\!\vct{a}_{2}}{r_{12}^3}+3m_{1}m_{2}
\frac{(\vct{a}_{1}\!\cdot\!\vct{n}_{12})(\vct{a}_{2}\!\cdot\!\vct{n}_{12})}{r_{12}^3}\;,
\end{aligned}
\end{align}

\begin{align}
\begin{aligned}
H_{S_{1}S_{2}^3}&=\frac{1}{12}m_{1}m_{2}\left[(\vct{a}_{1}\!\cdot\!\vct{a}_{2})
(\vct{\partial}_{1}\!\cdot\!\vct{\partial}_{2})^2\widehat{(\vct{a}_{2}\!\cdot\!\vct{\partial}_{2})^2}-(\vct{a}_{1}\!\cdot\!\vct{\partial}_{2})
(\vct{a}_{2}\!\cdot\!\vct{\partial}_{1})(\vct{\partial}_{1}\!\cdot\!\vct{\partial}_{2})\widehat{(\vct{a}_{2}\!\cdot\!\vct{\partial}_{2})^2}\right]r_{12}\\
&=-\frac{1}{12}m_{1}m_{2}(\vct{a}_{1}\!\cdot\!\vct{\partial}_{2})(\vct{a}_{2}\!\cdot\!\vct{\partial}_{1})(\vct{\partial}_{1}\!\cdot\!\vct{\partial}_{2})
\widehat{(\vct{a}_{2}\!\cdot\!\vct{\partial}_{2})^2}r_{12}\\
&=m_{1}m_{2}\bigg(-\frac{3}{2}\frac{(\vct{a}_{1}\!\cdot\!\vct{a}_{2})\vct{a}_{2}^2}{r_{12}^5}+\frac{15}{2}
\frac{(\vct{a}_{2}\!\cdot\!\vct{n}_{12})^2(\vct{a}_{1}\!\cdot\!\vct{a}_{2})}{r_{12}^5}+\frac{15}{2}
\frac{\vct{a}_{2}^2(\vct{a}_{1}\!\cdot\!\vct{n}_{12})(\vct{a}_{2}\!\cdot\!\vct{n}_{12})}{r_{12}^5}\\
&\quad\quad\quad\quad-\frac{35}{2}\frac{(\vct{a}_{1}\!\cdot\!\vct{n}_{12})(\vct{a}_{2}\!\cdot\!\vct{n}_{12})^3}{r_{12}^5}\bigg)\\
&=H_{S_2S_{1}^3}(1\leftrightarrow 2)\,.
\end{aligned}
\end{align}

We sum up expressions

\begin{align}
\begin{aligned}
H_{S_{1}^{2}} + H_{S_{2}^{2}} + H_{S_{1}S_{2}} =\;\frac{1}{2}\frac{m_{1}m_{2}}{r_{12}^{3}}\left(3\left(\vct{a}_{1}\!\cdot\!\vct{n}_{12}\right)^2+
3\left(\vct{a}_{2}\!\cdot\!\vct{n}_{12}\right)^2-\vct{a}_{1}^2-\vct{a}_{2}^2+2\vct{a}_{1}\!\cdot\!\vct{a}_{2}-
6\left(\vct{a}_{1}\!\cdot\!\vct{n}_{12}\right)\left(\vct{a}_{2}\!\cdot\!\vct{n}_{12}\right)\right)\,,
\end{aligned}
\end{align}

which is the same expression as in e.g., \cite{D01, schafermagnetic};
also see \cite{barker, TH85}.

\subsection{Linear in $G$ and $p$ spin-orbit interaction Hamiltonians}

Taking the partly boosted sources for Hamiltonian and
momentum constraint into consideration we will be able
to compute the spin-orbit Hamiltonian with some higher spin
corrections. First we look at the binary source of the Hamiltonian constraint

\begin{equation}
-\Delta\phi_{\text{I}}=-\Delta\phi_{\text{I}}^{(1)}-\Delta\phi_{\text{I}}^{(2)}
\end{equation}

with

\begin{align}
-\Delta\phi_{\text{I}}^{(1)}&:=16\pi\left(\left(1-\frac{1}{2}\widehat{(\vct{a}_{1}\!\cdot\!\vct{\partial}_{1})^2}\right)m_{1}+\frac{1}{2}\vct{p}_1\!\cdot\!\left(\vct{a}_{1}\times\vct{\partial}_{1}\right)\right)\delta_{1}+16\pi\left(\left(1-\frac{1}{2}\widehat{(\vct{a}_{2}\!\cdot\!\vct{\partial}_{2})^2}\right)m_{2}+\frac{1}{2}\vct{p}_2\!\cdot\!\left(\vct{a}_{2}\times\vct{\partial}_{2}\right)\right)\delta_{2}\label{lalala}\\
-\Delta\phi_{\text{I}}^{(2)}&:=\left(\pi^{ij}\right)^2\,.
\end{align}

Integration over $-\Delta\phi_{\text{I}}^{(1)}$ will not contribute to spin-orbit-interaction, but $-\Delta\phi_{\text{I}}^{(2)}$ will. So we now have to solve the momentum constraint.

\begin{equation}
\pi^{ij}_{,\,j}=\pi^{ij}_{1,\,j}+\pi^{ij}_{2,\,j}
\end{equation}
with
\begin{equation}
\pi^{ij}_{1,\,j}=-4\pi \epsilon^{ikl}a_{1k}\partial_{1l}\left(1-\frac{1}{6}\widehat{(\vct{a}_{1}\!\cdot\!\vct{\partial}_{1})^2}\right)m_{1}\delta_{1}
-8\pi p_{1}^{i}\delta_{1}=\pi^{ij}_{2,\,j}(2\leftrightarrow 1)\,
\end{equation}
which gives

\begin{align}
\begin{aligned}
\pi^{ij}_{1}&=\;-m_{1}\epsilon^{ipl}\partial^{j}_{1}a_{1p}\partial_{1l}\left(1-\frac{1}{6}\widehat{(\vct{a}_{1}\!\cdot\!\vct{\partial}_{1})^2}\right)\frac{1}{r_{1}}-m_{1}\epsilon^{jpl}\partial^{i}_{1}a_{1p}\partial_{1l}\left(1-\frac{1}{6}\widehat{(\vct{a}_{1}\!\cdot\!\vct{\partial}_{1})^2}\right)\frac{1}{r_{1}}\\
&\quad\;+p_{1k}\left(\delta_{ij}\partial_{1k}\frac{1}{r_{1}}-2\delta_{ik}\partial_{1j}\frac{1}{r_{1}}-2\delta_{jk}\partial_{1i}\frac{1}{r_{1}}+\frac{1}{2}\partial_{1i}\partial_{1j}\partial_{1k}r_{1}\right)=\pi^{ij}_{2}(2\leftrightarrow 1)\,
\end{aligned}
\end{align}

resulting in an interaction contribution (indicated by $\simeq$):

\begin{equation}
\begin{split}
2\pi^{ij}_{1}\pi^{ij}_{2}\simeq&\;2\bigg[-m_{1}\epsilon^{ipl}\partial^{j}_{1}a_{1p}\partial_{1l}\left(1-\frac{1}{6}\widehat{(\vct{a}_{1}\!\cdot\!\vct{\partial}_{1})^2}\right)\frac{1}{r_{1}}p_{2k}\left(-2\delta_{ik}\partial_{2j}\frac{1}{r_{2}}-2\delta_{jk}\partial_{2i}\frac{1}{r_{2}}+\frac{1}{2}\partial_{2i}\partial_{2j}\partial_{2k}r_{2}\right)\\
&\quad-m_{1}\epsilon^{jpl}\partial^{i}_{1}a_{1p}\partial_{1l}\left(1-\frac{1}{6}\widehat{(\vct{a}_{1}\!\cdot\!\vct{\partial}_{1})^2}\right)\frac{1}{r_{1}}p_{2k}\left(-2\delta_{ik}\partial_{2j}\frac{1}{r_{2}}-2\delta_{jk}\partial_{2i}\frac{1}{r_{2}}+\frac{1}{2}\partial_{2i}\partial_{2j}\partial_{2k}r_{2}\right)\\
&\quad-m_{2}\epsilon^{ipl}\partial^{j}_{2}a_{2p}\partial_{2l}\left(1-\frac{1}{6}\widehat{(\vct{a}_{2}\!\cdot\!\vct{\partial}_{2})^2}\right)\frac{1}{r_{2}}p_{1k}\left(-2\delta_{ik}\partial_{1j}\frac{1}{r_{1}}-2\delta_{jk}\partial_{1i}\frac{1}{r_{1}}+\frac{1}{2}\partial_{1i}\partial_{1j}\partial_{1k}r_{1}\right)\\
&\quad-m_{2}\epsilon^{jpl}\partial^{i}_{2}a_{2p}\partial_{2l}\left(1-\frac{1}{6}\widehat{(\vct{a}_{2}\!\cdot\!\vct{\partial}_{2})^2}\right)\frac{1}{r_{2}}p_{1k}\left(-2\delta_{ik}\partial_{1j}\frac{1}{r_{1}}-2\delta_{jk}\partial_{1i}\frac{1}{r_{1}}+\frac{1}{2}\partial_{1i}\partial_{1j}\partial_{1k}r_{1}\right)\bigg]
\end{split}
\end{equation}

and leading to a spin-orbit Hamiltonian

\begin{equation}
\begin{split}
H_{SO}^{(a)}=&\;\frac{1}{16\pi}\int-\Delta\phi_{\text{I}}^{(2)}\dd^{3}x\\
=&\;\frac{1}{16\pi}\int\left(\pi^{ij}\right)^{2}\dd^{3}x\\
\simeq&\;-m_{1}\epsilon^{ipl}p_{2i}a_{1p}\vct{\partial}_{1}\!\cdot\!\vct{\partial}_{2}\partial_{1l}\left(1-\frac{1}{6}\widehat{(\vct{a}_{1}\!\cdot\!\vct{\partial}_{1})^2}\right)r_{12}-m_{2}\epsilon^{ipl}p_{1i}a_{2p}\vct{\partial}_{1}\!\cdot\!\vct{\partial}_{2}\partial_{2l}\left(1-\frac{1}{6}\widehat{(\vct{a}_{2}\!\cdot\!\vct{\partial}_{2})^2}\right)r_{12}\\
=&\;-2m_{1}\epsilon^{ipl}p_{2i}a_{1p}\left(1-\frac{1}{6}\widehat{(\vct{a}_{1}\!\cdot\!\vct{\partial}_{1})^2}\right)\frac{n_{12l}}{r_{12}^{2}}+2m_{2}\epsilon^{ipl}p_{1i}a_{2p}\left(1-\frac{1}{6}\widehat{(\vct{a}_{2}\!\cdot\!\vct{\partial}_{2})^2}\right)\frac{n_{12l}}{r_{12}^{2}}\\
=&\;\vct{S}_{1}\!\cdot\!\left(\vct{n_{12}}\times\vct{p}_{2}\right)\left(-\frac{2}{r_{12}^2}-\frac{\vct{a}_{1}^2}{r_{12}^4}+\frac{5(\vct{a}_{1}\!\cdot\!\vct{n}_{12})^2}{r_{12}^4}\right)+\vct{S}_{2}\!\cdot\!\left(\vct{n_{12}}\times\vct{p}_{1}\right)\left(\frac{2}{r_{12}^2}+\frac{\vct{a}_{2}^2}{r_{12}^4}-\frac{5(\vct{a}_{2}\!\cdot\!\vct{n}_{12})^2}{r_{12}^4}\right)\,.
\end{split}
\end{equation}

For some further contributions we calculate 

\begin{equation}
\phi_{\text{I}}^{(1)}=4\left[\left(1-\frac{1}{2}\widehat{(\vct{a}_{1}\!\cdot\!\vct{\partial}_{1})^2}\right)\frac{m_{1}}{r_{1}}+\frac{1}{2}\vct{p}_{1}\!\left(\vct{a}_{1}\times\vct{\partial}_{1}\right)\frac{1}{r_{1}}\right]+4\left[\left(1-\frac{1}{2}\widehat{(\vct{a}_{2}\!\cdot\!\vct{\partial}_{2})^2}\right)\frac{m_{2}}{r_{2}}+\frac{1}{2}\vct{p}_{2}\!\left(\vct{a}_{2}\times\vct{\partial}_{2}\right)\frac{1}{r_{2}}\right]\,,
\end{equation}

which leads to the next order source to be integrated

\begin{equation}
\begin{split}
-\Delta\phi_{\text{II}}=&\;16\pi\left[-\frac{1}{8}\phi_{\text{I}}^{(1)}m_{1}\left(1-\widehat{\frac{1}{2}(\vct{a}_{1}\!\cdot\!\vct{\partial}_{1})^2}\right)-\frac{5}{16}\phi_{\text{I}}^{(1)}\vct{p}_{1}\!\cdot\!\left(\vct{a}_{1}\times\vct{\partial}_{1}\right)\right]\delta_{1}\\
&\;+16\pi\left[-\frac{1}{8}\phi_{\text{I}}^{(1)}m_{2}\left(1-\widehat{\frac{1}{2}(\vct{a}_{2}\!\cdot\!\vct{\partial}_{2})^2}\right)-\frac{5}{16}\phi_{\text{I}}^{(1)}\vct{p}_{2}\!\cdot\!\left(\vct{a}_{2}\times\vct{\partial}_{2}\right)\right]\delta_{2}\\\\
\simeq&\;\frac{16\pi}{4}\vct{p}_{2}\!\cdot\!\left(\vct{a}_{2}\times\vct{\partial}_{2}\right)\frac{1}{r_{2}}\left(-1+\widehat{\frac{1}{2}(\vct{a}_{1}\!\cdot\!\vct{\partial}_{1})^2}\right)m_{1}\delta_{1}+\frac{16\pi}{4}\vct{p}_{1}\!\cdot\!\left(\vct{a}_{1}\times\vct{\partial}_{1}\right)\frac{1}{r_{1}}\left(-1+\widehat{\frac{1}{2}(\vct{a}_{2}\!\cdot\!\vct{\partial}_{2})^2}\right)m_{2}\delta_{2}\\
&\;+16\pi\frac{5}{4}\vct{p}_{1}\!\cdot\!\left(\vct{a}_{1}\times\vct{\partial}_{1}\right)\delta_{1}\left(-1+\widehat{\frac{1}{2}(\vct{a}_{2}\!\cdot\!\vct{\partial}_{2})^2}\right)\frac{m_{2}}{r_{2}}+16\pi\frac{5}{4}\vct{p}_{2}\!\cdot\!\left(\vct{a}_{2}\times\vct{\partial}_{2}\right)\delta_{2}\left(-1+\widehat{\frac{1}{2}(\vct{a}_{1}\!\cdot\!\vct{\partial}_{1})^2}\right)\frac{m_{1}}{r_{1}}\,,
\end{split}
\end{equation}

giving rise to a Hamiltonian

\begin{equation}
\begin{split}
H_{SO}^{(b)}=&\;-\frac{1}{16\pi}\int\Delta\phi_{\text{II}}\dd^{3}x\\
\simeq&\;\frac{1}{4}\vct{p}_{2}\!\cdot\!\left(\vct{a}_{2}\times\vct{\partial}_{2}\right)\left(-1+\widehat{\frac{1}{2}(\vct{a}_{1}\!\cdot\!\vct{\partial}_{1})^2}\right)\frac{m_{1}}{r_{12}}+\frac{1}{4}\vct{p}_{1}\!\cdot\!\left(\vct{a}_{1}\times\vct{\partial}_{1}\right)\left(-1+\widehat{\frac{1}{2}(\vct{a}_{2}\!\cdot\!\vct{\partial}_{2})^2}\right)\frac{m_{2}}{r_{12}}\\
&\;+\frac{5}{4}\vct{p}_{1}\!\cdot\!\left(\vct{a}_{1}\times\vct{\partial}_{1}\right)\left(-1+\widehat{\frac{1}{2}(\vct{a}_{2}\!\cdot\!\vct{\partial}_{2})^2}\right)\frac{m_{2}}{r_{12}}+\frac{5}{4}\vct{p}_{2}\!\cdot\!\left(\vct{a}_{2}\times\vct{\partial}_{2}\right)\left(-1+\widehat{\frac{1}{2}(\vct{a}_{1}\!\cdot\!\vct{\partial}_{1})^2}\right)\frac{m_{1}}{r_{12}}\\
=&\;\frac{3}{2}\vct{p}_{2}\!\cdot\!\left(\vct{a}_{2}\times\vct{\partial}_{2}\right)\left(-1+\widehat{\frac{1}{2}(\vct{a}_{1}\!\cdot\!\vct{\partial}_{1})^2}\right)\frac{m_{1}}{r_{12}}+\frac{3}{2}\vct{p}_{1}\!\cdot\!\left(\vct{a}_{1}\times\vct{\partial}_{1}\right)\left(-1+\widehat{\frac{1}{2}(\vct{a}_{2}\!\cdot\!\vct{\partial}_{2})^2}\right)\frac{m_{2}}{r_{12}}\\
=&\;\frac{3}{2}\frac{m_{2}}{m_{1}}\left[\frac{\vct{p}_{1}\!\cdot\!\left(\vct{S}_{1}\times\vct{n}_{12}\right)}{r_{12}^{2}}+\frac{1}{2}\left(\frac{3\vct{a}_{2}^{2}\,\vct{p}_{1}\!\cdot\!\left(\vct{S}_{1}\times\vct{n}_{12}\right)}{r_{12}^4}+\frac{6(\vct{a}_{2}\!\cdot\!\vct{n}_{12})\,\vct{p}_{1}\!\cdot\!\left(\vct{S}_{1}\times\vct{a}_{2}\right)}{r_{12}^4}-\frac{15(\vct{a}_{2}\!\cdot\!\vct{n}_{12})^2\,\vct{p}_{1}\!\cdot\!\left(\vct{S}_{1}\times\vct{n}_{12}\right)}{r_{12}^4}\right)\right]\\
&\;-\frac{3}{2}\frac{m_{1}}{m_{2}}\left[\frac{\vct{p}_{2}\!\cdot\!\left(\vct{S}_{2}\times\vct{n}_{12}\right)}{r_{12}^{2}}+\frac{1}{2}\left(\frac{3\vct{a}_{1}^{2}\,\vct{p}_{2}\!\cdot\!\left(\vct{S}_{2}\times\vct{n}_{12}\right)}{r_{12}^4}+\frac{6(\vct{a}_{1}\!\cdot\!\vct{n}_{12})\,\vct{p}_{2}\!\cdot\!\left(\vct{S}_{2}\times\vct{a}_{1}\right)}{r_{12}^4}-\frac{15(\vct{a}_{1}\!\cdot\!\vct{n}_{12})^2\,\vct{p}_{2}\!\cdot\!\left(\vct{S}_{2}\times\vct{n}_{12}\right)}{r_{12}^4}\right)\right]\,.
\end{split}
\end{equation}

Finally the extended spin-orbit Hamiltonian reads, cf.,
e.g. \cite{barker, D01}, 

\begin{equation}
\begin{split}
H_{SO}^{(a)} + H_{SO}^{(b)} =&\;\vct{S}_{1}\!\cdot\!\left(\vct{n_{12}}\times\vct{p}_{2}\right)\left(-\frac{2}{r_{12}^2}-\frac{\vct{a}_{1}^2}{r_{12}^4}+\frac{5(\vct{a}_{1}\!\cdot\!\vct{n}_{12})^2}{r_{12}^4}\right)+\frac{3}{2}\frac{m_{2}}{m_{1}}\vct{S}_{1}\!\cdot\!\frac{\left(\vct{n}_{12}\times\vct{p}_{1}\right)}{r_{12}^{2}}\\
&\;+\frac{3}{4}\frac{m_{2}}{m_{1}}\left(\frac{3\vct{a}_{2}^{2}\,\vct{S}_{1}\!\cdot\!\left(\vct{n}_{12}\times\vct{p}_{1}\right)}{r_{12}^4}+\frac{6(\vct{a}_{2}\!\cdot\!\vct{n}_{12})\,\vct{S}_{1}\!\cdot\!\left(\vct{a}_{2}\times\vct{p}_{1}\right)}{r_{12}^4}-\frac{15(\vct{a}_{2}\!\cdot\!\vct{n}_{12})^2\,\vct{S}_{1}\!\cdot\!\left(\vct{n}_{12}\times\vct{p}_{1}\right)}{r_{12}^4}\right)\\
&\;+\vct{S}_{2}\!\cdot\!\left(\vct{n_{12}}\times\vct{p}_{1}\right)\left(\frac{2}{r_{12}^2}+\frac{\vct{a}_{2}^2}{r_{12}^4}-\frac{5(\vct{a}_{2}\!\cdot\!\vct{n}_{12})^2}{r_{12}^4}\right)-\frac{3}{2}\frac{m_{1}}{m_{2}}\vct{S}_{2}\!\cdot\!\frac{\left(\vct{n}_{12}\times\vct{p}_{2}\right)}{r_{12}^{2}}\\
&\;-\frac{3}{4}\frac{m_{1}}{m_{2}}\left(\frac{3\vct{a}_{1}^{2}\,\vct{S}_{2}\!\cdot\!\left(\vct{n}_{12}\times\vct{p}_{2}\right)}{r_{12}^4}+\frac{6(\vct{a}_{1}\!\cdot\!\vct{n}_{12})\,\vct{S}_{2}\!\cdot\!\left(\vct{a}_{1}\times\vct{p}_{2}\right)}{r_{12}^4}-\frac{15(\vct{a}_{1}\!\cdot\!\vct{n}_{12})^2\,\vct{S}_{2}\!\cdot\!\left(\vct{n}_{12}\times\vct{p}_{2}\right)}{r_{12}^4}\right)\,.
\end{split}
\end{equation}

\bigskip

We state that the total angular momentum
$\vct{J} =\vct{L}+\vct{S}_{1}+\vct{S}_{2}$, with spin vectors
$\vct{S}_{a}=m\vct{a}_{a}$ ($a = 1,2$) and orbital angular momentum
$\vct{L}= \sum_a \vct{x}_{a}\times\vct{p}_{a}$, is conserved in time, 

\begin{equation}
\frac{\dd\vct{J}}{\dd t}=\left\{\vct{J}\,,\, H\right\}=0\,,
\end{equation}
where $H = H_{S_1p_2} + H_{S_1p_1}+H_{S_2p_1}+H_{S_2p_2}+H_{S_1S_2}+ H_{S_1^2}+
H_{S_2^2} + H_{S_1^3p_2}+H_{S_2^3p_1}+H_{S_1^2S_{2}p_{2}}+H_{S_2^2S_{1}p_{1}}+H_{S_1^2S_2^2}+H_{S_1S_2^3}+H_{S_2S_1^3}$, 
by making use of the standard Poisson brackets
\begin{align}
\left\{x_a^{i},\,p_{bj}\right\}=\delta_{ij}\delta_{ab}\;,\quad\quad
\left\{S_{ai},\,S_{bj}\right\}=\epsilon_{ijk}S_{ak}\delta_{ab}\,, \quad\quad \mbox{and zero otherwise}\,.
\end{align}
Also conserved are the absolute values of the spin vectors,
i.e. $\vct{S}_{a}^2=$ const. This is consistent with results from the
literature \cite{H73, PW01}, where it is shown that the tidal friction
enters on the spin-interaction-separation scaling of $(1/r_{12})^6$ only.  
It is also consistent with Ref. \cite{D83} where tidal deformations
for black holes are identified to occur at 5th post-Newtonian order for the first
time, whereas our developments do not exceed the 4th post-Newtonian level.
Obviously, our gravitating mass parameters $m_a$ do not change in
relation to the irreducible mass parameters $M_a$ as given by
$m_a^2 = M^2_a+ \vct{S}_{a}^2/4 M^2_a$, see \cite{C70}.

\section{Conclusions}
\label{sec:discussion}

In this paper several new interaction Hamiltonians for spinning binary black
holes have been derived,
$H_{S_2^2S_1p_1} + H_{S_2^3p_1} + H_{S_1^3p_2} + H_{S_1^2S_2p_2},~ H_{S_1^2S_2^2},~
H_{S_1S_2^3}+H_{S_2S_1^3}$.
These Hamiltonians generalize the previously known ones,
$H_{S_1p_1}+H_{S_2p_1}+H_{S_1p_2}+H_{S_2p_2},~H_{S_1S_2},~H_{S_1^2}+H_{S_2^2}$,
through further nonlinear terms in the 
spins. If the spins are counted of the order $(1/c)^0$, the new
Hamiltonains are of the order $1/c^2$ higher compared with the
previously known ones. Other Hamiltonians which are of the same higher
order in $1/c$, i.e. $1/c^4$, read $H_{S_1p_2(p^2+G)}+H_{S_2p_1(p^2+G)},~
H_{S_1S_2(p^2+G)}, ~H_{S_1^2(p^2+G)}+H_{S_2^2(p^2+G)}, ~H_{S_{1}^2S_{2}p_{1}}+H_{S_{2}^2S_{1}p_{2}}, ~H_{p_{1}S_{1}^3}+H_{p_{2}S_{2}^3}, ~H_{S_1^4}+H_{S_2^4}$. The
Hamiltonian  $H_{S_1p_2(p^2+G)}+H_{S_2p_1(p^2+G)}$
has recently been found by Damour, Jaranowski, and Sch\"afer
\cite{DJS}, the Hamiltonian $H_{S_1S_2(p^2+G)}$ may possibly
result from a recent paper by Porto and Rothstein, \cite{PR}. The 
Hamiltonians $H_{S_1^2(p^2+G)}+H_{S_2^2(p^2+G)}, ~H_{S_{1}^2S_{2}p_{1}}+H_{S_{2}^2S_{1}p_{2}}, ~H_{p_{1}S_{1}^3}+H_{p_{2}S_{2}^3}$ and
$H_{S_1^4}+H_{S_2^4}$ are still unknown.
In forthcoming papers, expressions for all the mentioned Hamiltonians
are expected to be presented as well as applications of them within
the context of gravitational wave astronomy.

\acknowledgments 
This work is supported by the Deutsche Forschungsgemeinschaft (DFG) through
SFB/TR7 ``Gravitational Wave Astronomy''.

\appendix
\section{Coordinate transformation}
\label{appendix_ct}

We start with an infinitesimal coordinate transformation with
translation vectors $\xi^{i}$, e.g. \cite{weinberg}, 

\begin{equation}
x^{\alpha}(x'^{\beta})=x'^{\alpha}+\xi^{\alpha}(x'^{\beta})\,,
\end{equation}
whereby $x^{\alpha}$ are the old coordinates and $x'^{\alpha}$ are the new ones the metric tensor is expressed in. The general transformation formula together with Taylor expansion leaves us with an expanded transformation formula
\begin{align}
\begin{aligned}
g_{\mu\nu}'(x'^{\alpha})&=g_{\alpha\beta}(x^{\sigma})\fracpd{x^{\alpha}}{x'^{\mu}}\fracpd{x^{\beta}}{x'^{\nu}}\\
&=g_{\alpha\beta}(x^{\sigma})\left(\delta^{\alpha}_{\mu}+?\xi^{\alpha}_{,\,\mu}?\right)\left(\delta^{\beta}_{\nu}+?\xi^{\beta}_{,\,\nu}?\right)\\
&=\left(g_{\alpha\beta}(x'^{\sigma})+g_{\alpha\beta,\,\lambda}(x'^{\sigma})\xi^{\lambda}+\hdots\right)\left(\delta^{\alpha}_{\mu}\delta_{\nu}^{\beta}+\delta^{\beta}_{\nu}?\xi^{\alpha}_{,\,\mu}?+\delta^{\alpha}_{\mu}?\xi^{\beta}_{,\,\nu}?+?\xi^{\alpha}_{,\,\mu}?\,?\xi^{\beta}_{,\,\nu}?\right)\\
&=g_{\mu\nu}+g_{\mu\nu,\,\lambda}\,\xi^{\lambda}+g_{\alpha\nu}\,?\xi^{\alpha}_{,\,\mu}?+g_{\beta\mu}\,?\xi^{\beta}_{,\,\nu}?+g_{\alpha\beta}\,?\xi^{\alpha}_{,\,\mu}?\,?\xi^{\beta}_{,\,\nu}?+\hdots\;.
\end{aligned}
\end{align}

Under the assumption that the metric tensor is stationary we specialize this formula for the components:

$\gamma_{ij}=g_{ij}$ 
\begin{eqnarray}
g_{00}'&=&g_{00}+g_{00,\,i}\,\xi^{i}+\hdots\label{g00trafo}\\
g_{0i}'&=&g_{0i}+g_{0i,\,j}\xi^{j}+g_{\alpha 0}?\xi^{\alpha}_{,\,i}?+\hdots\label{g0itrafo}\\
\gamma_{ij}'&=&\gamma_{ij}+\gamma_{ki}?\xi^{k}_{,\,j}?+\gamma_{kj}?\xi^{k}_{,\,i}?+\gamma_{ij,\,k}\xi^{k}+\gamma_{kl}?\xi^{k}_{,\,i}??\xi^{l}_{,\,j}?\nonumber\\
& &+g_{0i}?\xi^{0}_{,\,j}?+g_{0j}?\xi^{0}_{,\,i}?+g_{00}?\xi^{0}_{,\,i}??\xi^{0}_{,\,j}?+g_{0k}?\xi^{0}_{,\,i}??\xi^{k}_{,\,j}?+g_{0k}?\xi^{k}_{,\,i}??\xi^{0}_{,\,j}?+\hdots\label{gijtrafo}
\end{eqnarray}

\section{Useful Formulas}

From the Ref. \cite{blanchet} we take the formula
\begin{equation}\label{formulalaplace}
\Delta(r^{\lambda}\hat{n}_{L})=(\lambda-l)(\lambda+l+1)r^{\lambda-2}\hat{n}_{L}\qquad(\forall\lambda\,\in\,\mathbbm{C}\,!)
\end{equation}
with radial distance $r$ and the short-cut notation $\hat{n}_{L}:=\widehat{n_{i_{1}}\ldots n_{i_{l}}}$, whereby $n^{i}=x^{i}/r$ is the normal unit vector and the hat indicates tracefreeness according to formulas
\begin{align}
\hat{n}_{L}&=\sum_{k=0}^{[l/2]}\;(-)^{k}\frac{(2l-2k-1)!!}{(2l-1)!!}\delta_{\lbrace i_{1}i_{2}}\ldots\delta_{i_{2k-1}i_{2k}}n_{i_{2k+1}\ldots i_{l}\rbrace}\,,\\
n_{L}&=\sum_{k=0}^{[l/2]}\;(-)^{k}\frac{(2l-4k+1)!!}{(2l-2k+1)!!}\delta_{\lbrace i_{1}i_{2}}\ldots\delta_{i_{2k-1}i_{2k}}\hat{n}_{i_{2k+1}\ldots i_{l}\rbrace}\,,
\end{align}
where $A_{\lbrace i_{1}\ldots i_{l}\rbrace}$ is the non-weighted sum $\sum_{\sigma\in S}A_{ i_{\sigma_{1}}\ldots i_{\sigma_{l}}}$ with $S$ as the smallest set of permutations $(1\ldots l)$ that fully symmetrizes $A_{\lbrace i_{1}\ldots i_{l}\rbrace}$ in $i_{1}\ldots i_{l}$.

Inverting (\ref{formulalaplace}) leads to the very useful formula
\begin{equation}
\Delta^{-1}\left(r^{\lambda} \hat{n}_{L}\right)=\frac{r^{\lambda+2}\hat{n}_{L}}{(\lambda+2-l)(\lambda+3+l)}\,.
\end{equation}

\section{TT-part of Kerr 3-metric in the leading order}
\label{appendix_tt}

We make an ansatz for the 3-metric that is to be TT-gauged
($b,c,d,e,f,g$ being constants)

\begin{equation}\label{mansatz}
\gamma_{ij}=b\frac{m^2a^2}{r^{\alpha}}\delta_{ij}+c\frac{m^2(\vct{a}\!\cdot\!\vct{n})^2}{r^{\alpha}}\delta_{ij}+d\frac{m^2a^2n_{i}n_{j}}{r^{\alpha}}+e\frac{m^2(\vct{a}\!\cdot\!\vct{n})^2n_{i}n_{j}}{r^{\alpha}}+2f\frac{m^2(\vct{a}\!\cdot\!\vct{n})a_{(i}n_{j)}}{r^{\alpha}}+g\frac{m^2a_{i}a_{j}}{r^{\alpha}}\;.
\end{equation}

An infinitesimal coordinate transformation is given by

\begin{equation}
\gamma_{ij}'=\gamma_{ij}+\xi_{i,\,j}+\xi_{j,\,i}
\end{equation}
with $\xi^{i}$ being set as 
\begin{equation}
\xi^{i}=A\frac{m^2a^2n^{i}}{r^{\alpha-1}}+B\frac{m^2(\vct{a}\!\cdot\!\vct{n})^2n^{i}}{r^{\alpha-1}}+
C\frac{m^2(\vct{a}\!\cdot\!\vct{n})a^{i}}{r^{\alpha-1}}\,,
\end{equation}
where  $A,\;B,\;C$ denote constants.

The transformed metric then becomes

\begin{align}
\begin{aligned}
\gamma_{ij}'&=(b+2A)\frac{m^2a^2}{r^{\alpha}}\delta_{ij}+(c+2B)\frac{m^2(\vct{a}\!\cdot\!\vct{n})^2}{r^{\alpha}}\delta_{ij}+(d-2\alpha A)\frac{m^2a^2n_{i}n_{j}}{r^{\alpha}}\\
&\quad+(e-2(\alpha+2)B)\frac{m^2(\vct{a}\!\cdot\!\vct{n})^2n_{i}n_{j}}{r^{\alpha}}+(2f+4B-2\alpha C)\frac{m^2(\vct{a}\!\cdot\!\vct{n})a_{(i}n_{j)}}{r^{\alpha}}+(g+2C)\frac{m^2a_{i}a_{j}}{r^{\alpha}}\,.
\end{aligned}
\end{align}
The coefficients will be chosen so that $\gamma_{ij}'$ gets TT-gauged.  Therefore, we use the condition
\begin{equation}
h_{ij,\,i}^{TT}=\left(\gamma_{ij}'-\frac{1}{3}\gamma_{kk}'\delta_{ij}\right)_{,\,i}\stackrel{!}{=}0\,.
\end{equation}

The coefficients then result in 

\begin{align}
A&=\frac{2d(\alpha-5)(\alpha-3)(\alpha+2)-(\alpha-6)(2e+(2+\alpha)(2f+g\alpha))}{4(\alpha-5)(\alpha-3)\alpha(\alpha+2)}\label{kofvec1}\,,\\
B&=-\frac{(2+\alpha)(2f+g\alpha)+2e(14+\alpha(\alpha-8))}{4(\alpha-5)(\alpha-3)(\alpha+2)}\label{kofvec2}\,,\\
C&=\frac{2e(6+(\alpha-6)\alpha)+(2+\alpha)(g\alpha(2\alpha-9)+2f(6+(\alpha-6)\alpha))}{2(\alpha-5)(\alpha-3)\alpha(2+\alpha)}\label{kofvec3}\,.
\end{align}

For $\alpha=4$ we find:

\begin{equation}
h_{ij}^{TT}=\frac{\eta}{6}\left[-\frac{m^2a^2}{r^4}\delta_{ij}+2\frac{m^2(\vct{a}\!\cdot\!\vct{n})^2}{r^4}\delta_{ij}+
2\frac{m^2a^2n_{i}n_{j}}{r^4}-6\frac{m^2(\vct{a}\!\cdot\!\vct{n})^2n_{i}n_{j}}{r^4}+\frac{m^2a_{i}a_{j}}{r^4}\right]
\end{equation}

with

\begin{equation}
\eta=e+6f+12g\,.
\end{equation}

\section{Methods of regularizations}

\subsection{Partie finie regularization according to Hadamard}

 Let us consider $f$ being a real function defined in an environment of the point $\vct{x}_{0}\;\in\;\mathbbm{R}^3$, except in this point where $f$ is singular. We define a family of complex-valued functions $f_{\vct{n}}$ as follows:
\begin{equation}
f_{\vct{n}}\;:\;\mathbbm{C}\ni\varepsilon\;\mapsto\;f_{\vct{n}}(\varepsilon):=f(\vct{x}_{0}+\varepsilon\vct{n})\;\in\;\mathbbm{C}\,.
\end{equation}
We expand $f_{\vct{n}}$ in a Laurent-series around $\varepsilon=0$:
\begin{equation}\label{entop}
f_{\vct{n}}(\varepsilon)=\sum^{\infty}_{m=-N}\;a_{m}(\vct{n})\varepsilon^{m}\,.
\end{equation}
The regularized value of $f$ at $\vct{x}_{0}$ is defined as the
coefficient at $\varepsilon^{0}$ in the expansion (\ref{entop})
mean-valued over all unit vectors $\vct{n}$, \cite{schaefer1985, jara, faye},
\begin{equation}
f_{reg}(\vct{x}_{0}):=\frac{1}{4\pi}\oint\dd\Omega\,a_{0}(\vct{n})\,.
\end{equation}
This formula can be used to calculate integrals with delta-distributions. We define
\begin{equation}
\int\dd^{3}x\,f(\vct{x})\delta(\vct{x}-\vct{x}_{a}):=f_{reg}(\vct{x}_{a})\,,
\end{equation}
which provides us with a formula for calculating Poisson integrals of the form
\begin{equation}\label{poissonreg}
\Delta^{-1}\left\lbrace\sum_{a}f(\vct{x})\delta_{a}\right\rbrace=\Delta^{-1}\left\lbrace\sum_{a}f_{reg}(\vct{x})\delta_{a}\right\rbrace=\sum_{a}f_{reg}(\vct{x}_{a})\Delta^{-1}\delta_{a}=-\frac{1}{4\pi}\sum_{a}f_{reg}(\vct{x}_{a})\frac{1}{r_{a}}\,.
\end{equation}
Two examples shall clarify this.
\begin{flalign}\label{b1}
(1)\qquad\Delta^{-1}\left(\frac{1}{r}\delta(\vct{x})\right)=-\frac{1}{4\pi}\left(\frac{1}{r}\right)^{reg}_{\vct{x}=0}\frac{1}{r}=0
&&
\end{flalign}
because
\begin{equation*}
f=\frac{1}{r}\quad\curvearrowright\quad
f_{\vct{n}}=f(0+\varepsilon\vct{n})=\frac{1}{\varepsilon}\quad\rightsquigarrow\quad\left(\frac{1}{r}\right)^{reg}_{\vct{x}=0}=0\,,
\end{equation*}
\\
\begin{flalign*}
(2)\qquad\Delta^{-1}\left(\frac{1}{r}\partial_{i}\partial_{j}\delta(\vct{x})\right)=0\,.&&
\end{flalign*}
To prove this relation we make use of the identity
\begin{align*}
\frac{1}{r}\partial_{i}\partial_{j}\delta(\vct{x})&=\partial_{i}\partial_{j}\left(\frac{1}{r}\delta(\vct{x})\right)-\delta(\vct{x})\partial_{i}\partial_{j}\frac{1}{r}-\partial_{i}\delta(\vct{x})\partial_{j}\frac{1}{r}-\partial_{j}\delta(\vct{x})\partial_{i}\frac{1}{r}\,,\\
\partial_{i}\delta(\vct{x})\partial_{j}\frac{1}{r}&=\partial_{i}\left(\delta(\vct{x})\partial_{j}\frac{1}{r}\right)-\delta(\vct{x})\partial_{i}\partial_{j}\frac{1}{r}\,,\\
\hookrightarrow\quad\frac{1}{r}\partial_{i}\partial_{j}\delta(\vct{x})&=\partial_{i}\partial_{j}\left(\frac{1}{r}\delta(\vct{x})\right)+\delta(\vct{x})\partial_{i}\partial_{j}\frac{1}{r}-\partial_{i}\left(\delta(\vct{x})\partial_{j}\frac{1}{r}\right)-\partial_{j}\left(\delta(\vct{x})\partial_{i}\frac{1}{r}\right)\,,
\end{align*}
and conclude
\begin{equation}\label{b2}
\begin{split}
\Delta^{-1}\left(\frac{1}{r}\partial_{i}\partial_{j}\delta(\vct{x})\right)=&\;\partial_{i}\partial_{j}\Delta^{-1}\left(\frac{1}{r}\delta(\vct{x})\right)+\Delta^{-1}\left(\delta(\vct{x})\partial_{i}\partial_{j}\frac{1}{r}\right)-\partial_{i}\Delta^{-1}\left(\delta(\vct{x})\partial_{j}\frac{1}{r}\right)-\partial_{j}\Delta^{-1}\left(\delta(\vct{x})\partial_{i}\frac{1}{r}\right)\\
\stackrel{(\ref{b1})}{=}&\;\partial_{i}\frac{1}{4\pi}\left(\frac{n_{j}}{r^2}\right)^{reg}_{\vct{x}=0}\frac{1}{r}+\partial_{j}\frac{1}{4\pi}\left(\frac{n_{i}}{r^2}\right)^{reg}_{\vct{x}=0}\frac{1}{r}-\frac{1}{4\pi}\left(\frac{n_{i}n_{j}-\delta_{ij}}{r^{3}}\right)^{reg}_{\vct{x}=0}=0\,,
\end{split}
\end{equation}
because of absence of an $\varepsilon^{0}$-coefficient.

\subsection{Riesz's formula}

From the Ref.  \cite{riesz} we adapt the famous formula
\begin{equation}
\left[\int\dd^{3}x\,r^{\alpha}_{1}r^{\beta}_{2}\right]_{reg}:=\pi^{3/2}\frac{\Gamma\left(\frac{\alpha+3}{2}\right)\Gamma\left(\frac{\beta+3}{2}\right)\Gamma\left(-\frac{\alpha+\beta+3}{2}\right)}{\Gamma\left(-\frac{\alpha}{2}\right)\Gamma\left(-\frac{\beta}{2}\right)\Gamma\left(\frac{\alpha+\beta+6}{2}\right)}r_{12}^{\alpha+\beta+3}\,,
\end{equation}
e.g., 
\begin{equation}\label{intf}
\left[\int\dd^{3}x\,\frac{1}{r_{a}r_{b}}\right]_{reg}=-2\pi r_{ab}\,.
\end{equation}

\end{document}